\documentclass[]{pasj01}
\jyear{2026}
\twocolumn 
\usepackage{booktabs}
\Received{$\langle$reception date$\rangle$}
\Accepted{$\langle$acception date$\rangle$}
\Published{$\langle$publication date$\rangle$}
\usepackage{url,lscape}

\definecolor{xlinkcolor}{rgb}{0,0,1}
\definecolor{xurlcolor}{cmyk}{0.78,0.17,0.09,0} 
\makeatletter
\let\old@ssect\@ssect 
\makeatother

\usepackage{amsmath}

\usepackage{natbib}
\usepackage{hyperref}

\makeatletter
\providecommand*{\theH@article}{0}
\renewcommand*{\theHchapter}{\if@in@appendix Alph\arabic{chapter}\else\arabic{chapter}\fi}

\renewcommand*{\theHfigure}{\theH@article.\arabic{figure}}
\renewcommand*{\theHtable}{\theH@article.\arabic{table}}

\makeatother

\usepackage[dvipsnames]{xcolor}
\hypersetup{citecolor=RoyalBlue}
\hypersetup{colorlinks=True, linkcolor=blue, citecolor=blue, urlcolor=xurlcolor}

\makeatletter
\def\@ssect#1#2#3#4#5#6{%
  \NR@gettitle{#6}
  \old@ssect{#1}{#2}{#3}{#4}{#5}{#6}
}
\makeatother
\usepackage[symbol]{footmisc}

\usepackage{tabularx}
\usepackage{threeparttablex}
\usepackage{booktabs}

\begin{document}

\title{Multiwavelength Probes of Cosmic Ray Transport in Molecular Cloud Structures}

\author{
Hayden P. H. \textsc{Ng}\altaffilmark{1,2 $\dag$},
Ellis R. \textsc{Owen}\altaffilmark{3,4 $\dag$ $\star$},
Naomi \textsc{Tsuji}\altaffilmark{5,6}, 
Szu-Ting \textsc{Chen}\altaffilmark{7,2}
}

\altaffiltext{1}{Department of Physics and Astronomy, University College London, Gower Street, London, WC1E 6BT, UK}
\altaffiltext{2}{Erlangen Centre for Astroparticle Physics (ECAP), Friedrich-Alexander-Universit\"{a}t Erlangen-N\"{u}rnberg, Nikolaus-Fiebiger-Str.\ 2, 91058 Erlangen, Germany}
\altaffiltext{3}{Astrophysical Big Bang Laboratory (ABBL), RIKEN Pioneering Research Institute (PRI), Wak\={o}, Saitama 351-0198, Japan}
\altaffiltext{4}{Theoretical Astrophysics, Department of Earth and Space Science, The University of Osaka, Toyonaka, Osaka 560-0043, Japan}
\altaffiltext{5}{Institute for Cosmic Ray Research, The University of Tokyo, Kashiwanoha 5-1-5, Kashiwa, Chiba 277-8582, Japan}
\altaffiltext{6}{RIKEN Center for Interdisciplinary Theoretical and Mathematical Sciences (iTHEMS), Wak\={o}, Saitama 351-0198, Japan}
\altaffiltext{7}{Institute of Astronomy, National Tsing Hua University, Hsinchu 30013, Taiwan\\\vspace{-0.3cm}}

\email{ellis.owen@riken.jp}

\altaffiltext{$\dag$}{\footnotesize These two authors contributed equally to this work.}


\maketitle

\begin{abstract}

We investigate how cosmic ray (CR) transport in molecular clouds and their substructures can be probed using multi-wavelength observations. 
The detailed microphysics regulating the penetration and coupling of CRs in dense molecular structures is unsettled. Self-generated turbulence can produce scattering and diffusive transport, while ion-neutral damping in cold, dense gas promotes ballistic CR propagation. 
We construct a self-consistent model framework for CR transport and interactions in magnetized molecular clouds, considering three limiting propagation scenarios: ballistic transport, diffusion, and a hybrid configuration featuring a diffusive envelope and quasi-ballistic core. By forward-modeling pion-decay $\gamma$-ray emissivities, CR-driven ionization-rate profiles, and electron synchrotron emission in the hard X-ray band, we connect GeV-scale attenuation and propagation signatures to independent diagnostics of secondary production and low-energy CR penetration. As an illustrative example, we apply our framework to the Taurus molecular cloud complex and selected embedded clumps. We show that CR scattering may be substantially enhanced on clump scales, with inferred CR diffusion coefficients suppressed relative to canonical interstellar medium (ISM) values at GeV energies. In this interpretation, CRs are more closely coupled with dense gas in the ISM, and 
a diffusive envelope boosts the effective gas column density encountered by the CRs. This increases the hadronic interaction rate in the cloud. In turn, the secondary CR electron injection is also increased, and  CR ionization rates are elevated at higher densities. We show that a hard X-ray synchrotron emission component is also generated, which may be detectable with near-future facilities. Finally, we discuss how
future $\gamma$-ray, X-ray, and ionization constraints will provide firm tests of CR propagation theories in molecular cloud environments. 
\end{abstract}

\section{Introduction}
\label{sec:introduction}

\footnotetext[0]{$^{\star}$ Initial stage of this work was conducted while E.R.O. was at the
Center for Informatics and Computation in Astronomy, National Tsing Hua University, Hsinchu 30013, Taiwan.}

The microphysics that regulates cosmic ray (CR) transport in complex, magnetized interstellar media remains unsettled. CR pressure gradients can drive the growth of magneto-sonic turbulence~\citep[e.g.][]{Skilling1976A&A, Cesarsky1978A&A, Ivlev2018ApJ, Phan2018MNRAS, Bustard2021ApJ}, causing self-modulation and diffusive transport, where CRs scatter efficiently off the self-excited waves~\citep[e.g.][]{Skilling1975MNRAS, Morlino2015MNRAS, Ivlev2018ApJ, Chernyshov2024PhRvD, Chernyshov2025PhRvD}. This can suppress the CR spectrum within interstellar clouds at energies that contribute most strongly to wave excitation~\citep[e.g.][]{Skilling1976A&A, Dogiel2018ApJ, Chernyshov2024PhRvD, Axen2024ApJ}. Conversely, in dense core regions of clouds, ion-neutral damping can suppress this turbulence~\citep{Lagage1983A&A, Felice2001ApJ, Draine2011piim}.  
This shifts transport toward the ballistic regime~\cite[e.g.][]{Ivlev2018ApJ, Dogiel2018ApJ}, where CRs effectively become decoupled from dense gas. Fast-transport prescriptions based on this decoupling are often adopted in simulation studies~\citep[e.g.][]{Farber2018ApJ, Armillotta2021ApJ, Habegger2024ApJ, Sike2025ApJ}. 

The interplay between these mechanisms across the hierarchical structure of the interstellar medium (ISM) sets how CRs interact with the different components of molecular cloud complexes, and determines their role in shaping the thermal~\citep{Guesten1985A&A, 2012ApJGlassgold, Ao2013A&A, Pazianotto2021ApJ, Owen2021ApJ, Brugaletta2025MNRAS}, ionization~\citep{Padovani2009A&A, Padovani2013A&A, Luo2024A&A, Washinoue2024ApJ} and chemical~\citep{Padovani2020SSRv, Indriolo2023ApJ, Phan2024MNRAS} configuration of dense clouds and even of their internal substructures (see also~\citealt{Strong2007ARNPS, Grenier2015ARA&A, Klessen2016SAAS, Owen2023Galax} for reviews). 

The effects of CRs in molecular clouds can be observed in multiple ways. The ionizing impact of sub-GeV CRs in dense molecular cores has long been studied~\citep[e.g.][]{Hayakawa1961PASJ, Spitzer1968ApJ}. 
Astrochemical tracers are commonly used to constrain CR penetration into cloud interiors. Observations have shown declining ionization rates with increasing column density~\citep{Bacalla2019A&A, Sabatini2023ApJ, Pineda2024A&A}, consistent with attenuation by ionization losses~\citep{Padovani2009A&A, Padovani2020SSRv}, magnetic mirroring~\citep{Desch2004ApJ} and/or self-modulation~\citep{Ivlev2018ApJ, Dogiel2018ApJ}. 
Recently developed diagnostics that are sensitive to different CR energies provide additional information. These include $\gamma$-ray nuclear de-excitation lines~\citep{Shi2024A&A}, Fe K$\alpha$ X-rays at 6.40 keV~\citep{Nobukawa2018ApJ, Shimaguchi2022PASJ, Rogers2022ApJ, Shi2024A&A}, $\gtrsim$GeV $\gamma$-rays from neutral pion decays~\citep[e.g.][]{Phan2020A&A, Aharonian2020PhRvD, Tsuji2025arXiv}, synchrotron emission from CR electrons in radio~\citep{Protheroe2008MNRAS, Rodriguez2013ApJ} and X-rays~\citep[e.g.][]{Tsuji2024ApJ}, and absorption/emission signatures in the sub-mm band of the ionizing impact driven by sub-GeV CR protons and electrons~\citep[e.g.][]{Caselli1998ApJ, Neufeld2017ApJ}. Multimessenger approaches, particularly in Galactic Center clouds, may also offer relatively clean probes of CR-gas interactions~\citep{Icecube2023Sci, Lai2026PhRvD} and hold potential for future tests of CR propagation physics in phase-complex media~\citep{Roy2023JCAP, Torres2025Univ}. 

CR effects have been measured on the scale of individual clouds using these techniques, including in extreme environments such as the Galactic Center~\citep{YusefZadeh2013ApJ, Rogers2022ApJ}, near supernova remnants~\citep{Zhong2023MNRAS, Yeung2023PASJ, Indriolo2023ApJ}, and even in phase-selected ISM components of external galaxies out to cosmic noon~\citep{Indriolo2018ApJ}. However, all of these observables are indirect and have important limitations. These include uncertainties in astrochemical models 
in the interpretation of ionization signatures, 
especially under physical conditions that deviate from those of the local ISM~\citep{Wakelam2005A&A, Albertsson2018ApJ, Luo2024A&A}, line-of-sight projection effects~\citep{Beaumont2013ApJ}, inconsistencies in gas density determinations~\citep{Abrahams2017ApJ, Lai2026PhRvD}, and uncertainties in magnetic field estimates, particularly on small spatial scales~\cite[see][for reviews]{Crutcher2012ARA&A, Li2021Galax, Pattle2023ASPC}. Systematic variation in cloud structure and CR flux, such as trends with galacto-centric radius, further complicate the interpretation of these measurements~\citep{RomanDuval2010ApJ, Rigby2019A&A, Aharonian2020PhRvD}.

These issues can be mitigated by focusing on individual molecular clouds. This avoids the added complexity introduced by large-scale ISM environmental trends that affect conclusions drawn from population or stacking analyses. Higher-energy CRs provide an additional advantage, as they are less influenced by low-energy modulation~\citep{Gleeson1968ApJ} or small-scale magnetic field variations~\citep{Padovani2011A&A, Seta2018MNRAS}. While CRs above a few hundred MeV contribute little to ionization~\citep[e.g.][]{Padovani2009A&A}, they do produce $\gamma$-rays through neutral pion decay in hadronic proton-proton (pp) collisions. Such $\gamma$-ray observations have been used to infer the CR spectrum in nearby molecular clouds~\citep[e.g.][]{Yang2014A&A, Neronov2017A&A, Mizuno2022ApJ}
and in Galactic Center cloud complexes~\citep[e.g.][]{Yang2015A&A}. 
Although resolving sub-cloud structures remains challenging due to instrumental limitations, recent $\gamma$-ray studies have revealed spatial variations in the CR spectrum across different regions of the same cloud~\citep{Yang2023NatAs, 2024A&AKamalYoussef, Zeng2024RAA, Jiang2025MNRAS}. This opens up nearby molecular clouds as laboratories for precision tests of CR propagation. 

In this work, we explore how well such clouds can be used to constrain CR transport physics. 
We put focus on future multi-wavelength studies that are sensitive to a broad range of CR energies, and to the interactions they undergo in diffuse and dense gas. 
This paper is arranged as follows. Section~\ref{sec:cr_clouds} outlines the relevant CR transport and interaction processes, and their observable signatures. In Section~\ref{sec:model}, we describe our molecular cloud model, including the density and magnetic field profiles used to compute CR propagation/interaction processes. Section~\ref{sec:results} presents 
our results, including an illustrative application of our model to the Taurus molecular cloud complex. 
We then discuss the implications for probing CR transport physics in the multi-phase ISM in Section~\ref{sec:discussion}. We provide a summary and present our conclusions in Section~\ref{sec:summary}.

\section{CRs in molecular clouds}
\label{sec:cr_clouds}

\subsection{CR transport and interaction processes}
\label{sec:cr_physics}

A simplified macroscopic description of relativistic CR transport that retains the key physical processes can be written as: 
\begin{align}
\frac{\partial n_i}{\partial t}
\;+\;\nabla\!\cdot\mathbf{F}_i
\;+\;\frac{\partial}{\partial E_i}\!\left[b_i (E_i, r)\,n_i\right]
\;=\;\Gamma_i(E_i, r)\;-\;\Lambda_i(E_i, r)\,n_i \, , 
\label{eq:transport_flux_form}
\end{align} 
where $n_i = n_i(E_i, r)$ is the differential number density of CRs of species $i$, at position $r$, and where total energy $E_i = \gamma_{i} m_i c^2$, where $\gamma_i$ is the species-specific Lorentz factor, $c$ is the speed of light and $m_i$ is the rest mass of the species. 
This 
formulation evolves the flux for an inward-propagating CR population. The terms in 
Equation~\ref{eq:transport_flux_form}, from left to right, represent: (1) the time evolution of the CR population, (2) the 
radial CR flux gradient, (3) continuous energy losses or gains, where $b_{i}(E_i) = \dot{E_i}$ (such that $b_{i}(E_i)<0$) denotes cooling processes specific to each species, (4) injection of CRs at a volumetric rate $\Gamma_i(E_i)$, and (5) catastrophic losses $\Lambda_i(E_i)$, accounting for attenuation via processes such as inelastic collisions or decay, with species-dependent loss rates. In a molecular cloud, the evolution of the CR ensemble occurs on timescales much shorter than the dynamical timescale. 
If neglecting re-acceleration and negligible effects of advection by bulk flows, the CR distribution can therefore be well described using a quasi-steady  approximation.

In our approach, the CR transport physics in Equation~\ref{eq:transport_flux_form} is specified by the flux gradient term (2). For ballistic propagation (free streaming without scattering), we adopt the inward flux from the external CR sea. 
In the absence of scattering, this is $\mathbf{F}_i =-\,\beta_{i}\,c\,n_i\,\hat{\mathbf r}$, where $\hat{\mathbf r}$ is the outward radial unit vector, and $\beta_{i} = (1-\gamma_{i}^{-2})^{1/2}$ is the particle velocity normalized to the speed of light $c$. For an isotropic external CR intensity 
$J_i(E_i)$ at the outer boundary (located at $r=R$), the inward-going boundary condition is $n_i^-(E_i, R) = \pi J_i(E_i)/\beta_{i} c$, since the inward hemispheric crossing flux from the surrounding isotropic CR sea is $\pi J_i(E_i)$. 
For purely diffusive propagation, the flux is 
$ \mathbf{F}_i = -D(E_i,r)\,\nabla n_i$, and Equation~\ref{eq:transport_flux_form} reduces to the standard diffusion-loss equation. 
In this situation, spatial diffusion would be governed by the energy- and position-dependent diffusion coefficient $D(E_i, r)$, which accounts for random-walk scattering in magnetic turbulence.\footnote{In our treatment, magnetic mirroring and focusing~\citep{Desch2004ApJ} are not explicitly included. They are expected to approximately cancel out in the single-peaked magnetic field configurations typical of molecular clouds~\citep{Silsbee2018ApJ}. These effects 
can modify the CR distribution 
by a factor of order unity in such environments~\citep{Desch2004ApJ, Owen2021ApJ}.} We parameterize this in the form: 
\begin{equation}
D(E_i, r) = D_0 \left[ \frac{r_L(E_i,\langle |B| \rangle)|_{r}}{r_{L,0}}\right]^{\delta} 
\label{eq:diff_coeff}
\end{equation} 
where $\langle |B| \rangle\vert_{r} = |B(r)|$ is the characteristic mean strength of the magnetic field at position $r$, and $r_L$ is the CR gyro-radius. The normalization constant $D_0 = 3.0\times 10^{28} \, \mathrm{cm}^2 \mathrm{s}^{-1}$ corresponds to the empirically inferred diffusion coefficient in the Milky Way ISM for a $1 \, \mathrm{GeV}$ CR in a $5 \, \mu \mathrm{G}$ magnetic field, with reference gyro-radius $r_{L,0}$~\cite[see][]{Berezinskii1990book}. 
The index $\delta$ encodes the spectral slope of magnetic turbulence. We adopt a representative value of 1/3 throughout this work. This corresponds to Kolmogorov turbulence~\citep[e.g.][]{Berezinskii1990book, Strong2007ARNPS}, but we note 
that our choice of assumed turbulence spectrum does not qualitatively affect our conclusions.  

\subsubsection{Protons}
\label{sec:protons}

Equation~\ref{eq:transport_flux_form} can be simplified depending on the CR species under consideration, as not all terms are relevant to the microphysics of each particle type. For CR protons, the dominant energy loss process in molecular cloud environments are Coulomb and ionization losses, and streaming losses resulting from the excitation of Alfv\'{e}nic turbulence, while radiative losses are negligible. Coulomb and ionization losses become important below a few 100 MeV, but they are only significant in dense regions with large column densities. In a molecular cloud with a high neutral gas fraction, ionization losses are more severe than Coulomb losses. 
We therefore approximate the total cooling rate as the 
sum of ionization and streaming losses, 
$b_{\rm p} = b_{\rm p}^{\rm ion}(E_{\rm p})+b_{\rm p}^{\rm str}(E_{\rm p})$, where streaming losses are only active in regions where CRs are propagating diffusively. The ionization loss rate is obtained from the proton energy loss function provided by \cite{Padovani2018A&A}, and the Alfv\'{e}nic streaming loss term is given by 
\begin{equation}
b_{\mathrm{p}}^{\rm{str}}(E_{\rm{p}}) 
    = \frac{E_{\rm{p}}}{\epsilon_{\mathrm{CR}}} H_{\rm{str}}, 
\end{equation}
where $\epsilon_{\rm{CR}}$ is the CR energy density and $H_{\rm{str}} = - v_{A}\, \nabla P_{\mathrm{CR}}$ is the CR streaming energy transfer rate per unit volume. $\nabla P_{\rm{CR}} = \mathrm{d} P_{\mathrm{CR}}/\mathrm{d}r$ is the CR pressure gradient along the radial direction, and $v_{A}$~($=|{\boldsymbol v}_{A}|)= B/\sqrt{4\pi \rho}$ is the Alfv\'en speed where $\rho = \sum_i\;\!n_{i}^{\rm{ion}} \;\! m_i$ is the mass density of the ions in the medium. 

At higher CR energies, protons can also undergo hadronic 
 pp interactions. This process becomes relevant above a threshold energy of $\sim$ 0.28 GeV, corresponding to the proton energy needed to produce a neutral pion through pp$\rightarrow$pp$\pi^0$. These pp interactions may be regarded as catastrophic losses, so we model them as an absorption process. This is because the particles are either physically destroyed and converted to secondary products in an interaction event, or the primary CR proton loses a large fraction of its energy in a single hadronic interaction. These interactions occur at a rate $\Lambda_{\rm p}(E_{\rm p}) = n_{\rm H} \sigma_{\rm pp}(E_{\rm p}) c$, where $\sigma_{\rm pp}(E_{\rm p})$ is the inelastic pp interaction cross-section~\cite[see][]{Kafexhiu2014PRD}, and $n_{\rm H}$ is the local gas density.

In the absence of any source of CR protons within the cloud, such that CRs only propagate in from the outer boundary,\footnote{This may not hold in some systems that can sustain in situ acceleration, e.g. in proto-stars/proto-stellar jets~\citep{Padovani2015A&A, Cecere2016ApJ, Padovani2016A&A} {or accreting black holes~\citep[e.g.][]{Rice2021ApJ, Kimura2025ApJ}}.} a simplified form of Equation~\ref{eq:transport_flux_form} for CR protons accounting for all relevant processes above, can be written as: 
\begin{equation}
\nabla\!\cdot\mathbf{F}_{\rm p} + \frac{\partial}{\partial E_{\rm p}} \left[b_{\rm p}(E_{\rm p}) \;\! n_{\rm p}\right] + \Lambda_{\rm p}(E_{\rm p}) n_{\rm p} = 0 \ , 
\label{eq:transport_equation_protons}
\end{equation} 
where $n_{\rm p}$ is the CR proton differential number density.

\subsubsection{Electrons}
\label{sec:electrons}

CR electrons are subject to diffusion and lose energy to ionization, bremsstrahlung and radiative cooling. In typical molecular cloud environments, radiative losses are synchrotron-dominated. 
Unlike protons, streaming losses of electrons are negligible. This is because they are much lighter than protons, so they follow the magnetic field closely and do not efficiently excite Alfv\'{e}n waves via the same streaming instability mechanism. We model CR electron cooling using the function $b_e = b_\mathrm{e}^{\rm ion}(E_{\mathrm{e}}) \, + b_\mathrm{e}^{\rm brem}(E_{\mathrm{e}}) + \, b_{\mathrm{e}}^{\rm syn}(E_{\mathrm{e}})$, where ionization and 
bremsstrahlung losses ($b_\mathrm{e}^{\rm ion}(E_{\mathrm{e}}) \, + b_\mathrm{e}^{\rm brem}(E_{\mathrm{e}})$) are together derived from 
the loss function provided by \citet{Padovani2018A&A}. The synchrotron cooling rate $b_{\mathrm{e}}^{\rm{syn}}(E_{\mathrm{e}})$ is given by 
\begin{equation}
    b_{\mathrm{e}}^{\rm{syn}}(E_{\mathrm{e}}) = -\frac{4}{3}\sigma_{T}c\gamma_{\mathrm{e}}^{2}U_B,
\end{equation}
where the Lorentz factor $\gamma_{\mathrm{e}} = E_{\mathrm{e}}/m_{\mathrm{e}}c^2$, and the magnetic energy density $U_{B} = B^{2}/8\pi$. 

Unlike protons, CR electrons do not undergo catastrophic interactions. 
However, they are formed as secondary products in the pp collisions that attenuate protons in dense gas. The pp process 
injects electrons and positrons by the decay of charged pions. 
This decay process is practically instantaneous after a pp interaction arises. 
It occurs over a timescale of $2.6\times 10^{-8}\;\!{\rm s}$~\citep{Patrignani2016ChPh}. In this decay, around 3/4 of the CR proton's energy is inherited by neutrinos, which we do not consider further in our calculations. The remainder is transferred to the secondary electrons or positrons. Hereafter both CR electrons and positrons are referred to as `electrons' for convenience. This is because their astrophysical effects in a molecular cloud are practically indistinguishable.

Under these considerations, the transport equation for CR electrons reduces to: 
\begin{equation}
\nabla\!\cdot\mathbf{F}_{\rm e} + \frac{\partial}{\partial E_{\rm e}} \left[b_e (E_{\rm e}) \;\! n_{\rm e}\right] =  \Gamma_{\rm e}^{\rm sec}(E_{\rm e}) \ ,
\label{eq:transport_equation_electrons}
\end{equation} 
where $n_{\rm e}$ is the differential number density of CR electrons. 

\subsection{Observable signatures of CR effects}
\label{sec:obs_signatures}

\subsubsection{$\gamma$-ray emission}
\label{sec:gamma_rays}

The pp interactions that inject secondary electrons (see Section~\ref{sec:electrons}) also produce $\gamma$-rays via the decay of neutral pions. These neutral pions decay rapidly via the electromagnetic channel ($\pi^{0} \rightarrow 2\gamma$) with a branching ratio of approximately 98.9 per cent. This decay occurs on a timescale of $2.6\times 10^{-17}$ s~\citep{2020PTEP}, meaning that $\gamma$-ray emission follows almost instantaneously after a pp interaction. 
Provided that foreground/background contamination or obscured sources do not substantially contribute to the observed $\gamma$-ray flux from a cloud, its $\gamma$-ray spectrum can serve as a clear probe of the underlying high-energy CR proton spectrum and spatial distribution, offering indirect but relatively strong tests of CR transport models. 

\subsubsection{X-ray emission}
\label{sec:x_rays}

Secondary electrons produced in pp interactions cool by emitting synchrotron radiation in the magnetic fields of dense cloud regions. If the parent CR proton energies are sufficiently high, the resulting secondary electrons can reach energies capable of producing X-ray photons through this process. Several dedicated studies have searched for such a secondary synchrotron component in high-energy environments likely to be strongly irradiated by CR protons, such as in clouds around supernova remnants~\citep{Huang2020MNRAS, Higurashi2020ApJ, Tsuji2024ApJ}. Although such studies have so far only resulted in non-detections, 
this emission may be within reach of future hard X-ray missions. 

\subsubsection{Ionization signatures}
\label{sec:ionization}

Ionization of the dense gas in molecular clouds is dominated by lower-energy CRs. This provides a different window on the low energy regime of the CR spectrum that complements high energy tracers (X-rays and $\gamma$-rays). The ionization cross-section for CR protons peaks around 100 MeV, and for electrons around 100 eV~\cite[e.g.][]{Padovani2009A&A}. The ionizing effects of CRs are therefore a key diagnostic of CR physics in this lower energy regime. Low-energy CRs are more susceptible to magnetic deflection and scattering, and their propagation is more sensitive to the magnetic field structure and strength within the cloud. Ionization-based diagnostics therefore offer complementary information to that provided by $\gamma$-ray and X-ray observations, probing different aspects of both CR transport and the cloud environment.  

The CR ionization rate itself is not directly observable. Instead, CRs initiate chemical reactions within clouds, beginning with the ionization of molecular hydrogen to form ${\rm H}_2^{+}$~\citep{Dalgarno2006PNAS}. Ion-neutral chemistry then follows rapidly via ${\rm H}_2^{+} + {\rm H}_2 \rightarrow {\rm H}_3^{+} + {\rm H}$, which initiates a network of reactions producing a range of hydrogenated ions. Observable products of this chemistry include species such as 
${\rm HCO}^+$, ${\rm N}_2{\rm H}^+$, ${\rm OH}^+$, ${\rm H}_2{\rm O}^+$ and ${\rm H}_3{\rm O}^+$. These have, among other species, become established as probes of the low-energy CR ionization rate in interstellar environments~\citep[e.g.][]{Caselli1998ApJ, Hollenbach2012ApJ, Ceccarelli2014ApJ, Indriolo2018ApJ, Gaches2019ApJ}. \footnote{Although detailed modeling of these astrochemical networks is beyond the scope of this work, we later present our results in terms of CR ionization rate profiles. These can be interfaced as inputs to dedicated astrochemical codes (e.g. {\tt UCLCHEM}, {\tt Astrochem} or {\tt 3D-PDR}; see~\citealt{Bisbas2012MNRAS, Maret2015ascl, Holdship2017AJ, Gaches2019ApJ}), {or used in post-processing approaches that connect CR propagation, attenuation and chemical evolution in dense clouds \citep[e.g.][]{Latrille2025A&A}.}}

\section{Model}
\label{sec:model}

\subsection{Physical properties of molecular clouds}
\label{sec:general_intro}

Molecular clouds are dense associations of interstellar gas, in pressure equilibrium with the hot ionized component of the ISM. They typically extend over a few to several tens of pc and have a hierarchical internal density structure. While the complexity of cloud substructure cannot always be captured by a simple discrete hierarchy, and various classification schemes exist~\citep[see][]{Rodriguez2005ASPC}, a rough distinction can be made. On intermediate scales, clouds contain clumps (or filamentary structures; see~\citealt{Arzoumanian2011A&A}) with sizes of $0.3\text{--}3$ pc and typical densities in the range $10^3$--$10^4$~cm$^{-3}$. These, in turn, enclose smaller-scale cores of 0.03--0.2 pc with higher densities of $10^4\text{--}10^5$~cm$^{-3}$, and in some cases reaching up to $10^6\text{--}10^7$ cm$^{-3}$~\citep[see][for a review]{Bergin2007ARAA}. The ionization fraction within clouds generally decreases with increasing density, ranging from nearly unity in the diffuse outer regions to trace levels as low as $\sim 10^{-9}$ reported in some dense cores that are strongly shielded from ionization agents~\citep[e.g.][]{Goicoechea2009A&A}. Magnetic fields 
suffuse the cloud structure, 
and generally strengthen toward denser regions~\citep{Crutcher2012ARA&A}, reaching up to a few tens of $\mu$G or more. In some cases, these fields may become dynamically significant, providing support against gravitational collapse alongside turbulence~\citep[see][for a review]{Pattle2023ASPC}, particularly when the ionization fraction is high enough to maintain coupling between the magnetic field and the cloud medium.

To capture these physical characteristics of molecular clouds, we construct a simplified molecular cloud model as a basis to investigate how different CR propagation scenarios affect observable CR signatures. 
As a baseline, we consider a spherically symmetric toy model cloud 
with a simple cored density profile
following 
\begin{equation}
    n_{\rm{H}}(r) = n_0 \left[a+ \left(\frac{r}{R_{\rm eff}}\right)^{2}\right]^{-1} \ ,
    \label{eq:n_gas}
\end{equation}
corresponding to a finite-density core of radius $\sqrt{a} R_{\rm eff}$, an outer
$n_{\rm{H}} \propto r^{-2}$ envelope, and an effective size $R_{\rm eff}$. Hereafter, we set $a = 0.05$, which gives a core radius of around $0.22 R_{\rm eff}$, roughly reflective of the 
flat inner region found in 
observed clumps which have sizes of around $10\text{--}30$ percent of the 
characteristic radius~\citep{Lin2022A&A, Hoang2025A&A}.
This profile captures the qualitative central flattening seen in many clumps~\cite[e.g.][]{Beuther2002ApJ, Beuther2024A&A} 
while retaining a standard isothermal-like outer slope, and avoids a central divergence. The cloud is 
spherically symmetric and truncated at a 
characteristic radius of $R = R_{\rm eff}$ to ensure a finite mass. Here, 
where $n_0$ is the density of the gas at the cloud center, which we leave as a model parameter to be specified. 
 We set the magnetic field strength to scale with 
the cloud density, according to the empirical relation obtained by \cite{Crutcher2010ApJ}:
\begin{equation}
  B(n_{\rm{H}}) = \begin{cases}
B_{\rm ISM} \ ,  \hspace*{2cm} n_{\rm{H}} \leq n_1 \ ;  \\
B_{\rm ISM}\left({{n_{\rm{H}}}/{n_1}}\right)^{\bar{q}} \ , \hspace*{0.47cm} \ \ n_{\rm{H}} > n_1  \ , 
\end{cases}    
\label{eq:crutcher_model}
\end{equation}
where 
$n_1 = 300~\text{cm}^{-3}$, $\bar{q} \approx 0.65$, 
and 
$B_{\rm ISM} = 10~\mu\text{G}$ is the background magnetic field strength appropriate for the ISM average. 

\begin{figure}
    \centering
    \includegraphics[width=0.9\linewidth]{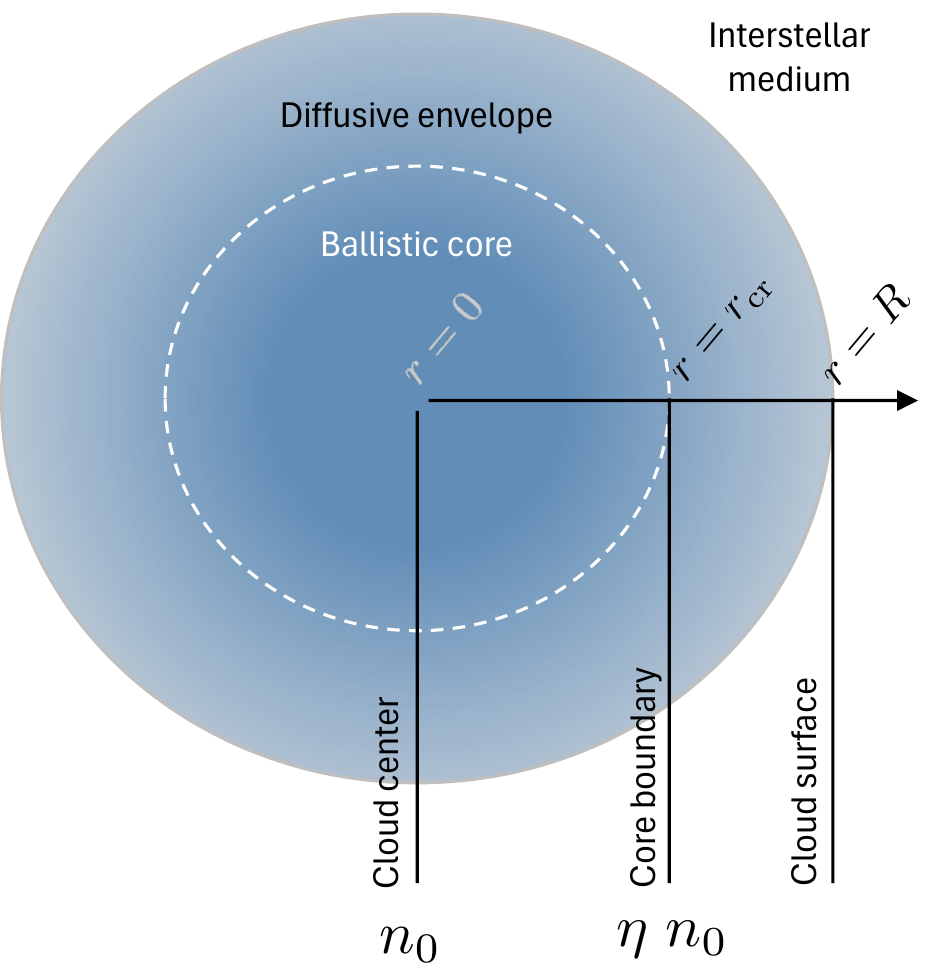}
    \caption{Schematic of our molecular cloud model and the generalized framework for model-agnostic, parameterized tests of CR transport. The color scale represents the cloud’s density structure. CR propagation regimes are shown as an outer diffusive envelope and an inner ballistic core, separated at a critical radius $r= r_{\rm cr}$ where the density exceeds a parameterized threshold $n_{\rm H}^{\rm c}=\eta \;\! n_0$. The central density at $r=0$ is denoted $n_0$  and the cloud boundary is set at $r=R$, which defines the system’s outer extent.}
    \label{fig:cloud_framework}
\end{figure}

Using this cloud model as a basis, we aim to construct a framework that is agnostic to microphysical models of CR transport, but demonstrates how different CR propagation models could manifest in observable signatures. To achieve this, we include three generic representative transport configurations in the context of our cloud model that allow a range of CR propagation theories to be tested. The first is a purely diffusive case, where CRs maintain diffusive behavior throughout the entire molecular cloud. The second is a purely ballistic case, in which CRs traverse the cloud without undergoing any scattering. The third configuration involves a transition: CRs diffuse through an outer envelope but switch to ballistic propagation in the dense core. This change in transport regime occurs at a {transition} density, parameterised in terms of a fraction of the central gas mass density $\eta n_0$ (cf. the clump in diffusive envelope  configuration discussed  by~\citealt{Chernyshov2024PhRvD, Chernyshov2025PhRvD}). The overall setup is illustrated in Figure~\ref{fig:cloud_framework}. 
In regions where propagation is dominated by 
diffusion, the flux term in Equation~\ref{eq:transport_flux_form} takes the form $ \mathbf{F}_i = -D(E_i,r)\,\nabla n_i$. In ballistic propagation regions, it instead is set to  $\mathbf{F}_i =-\,\beta_{i}(\gamma_{i})\,c\,n_i\,\hat{\mathbf r}$. 

 \begin{table*}
  \centering
  \setlength{\tabcolsep}{6pt}
  \renewcommand{\arraystretch}{1.15}
  \begin{tabular}{llll}
    \toprule
    \textbf{Parameter} & \textbf{Definition \& notes} & \textbf{Cloud} & \textbf{Clump} \\
    \midrule
    $d$ & Distance$^{\,a}$ & $150\,\mathrm{pc}$ & $150\,\mathrm{pc}$ \\
    $R_{\rm eff}$ & Effective radius$^{\,b}$ & $30\,\mathrm{pc}$ & $1\,\mathrm{pc}$ \\
    $\langle n_{\rm H}\rangle$ & Mean hydrogen density$^{\,b}$ & $3.0\times10^{1}\,\mathrm{cm^{-3}}$ & $3.0\times10^{4}\,\mathrm{cm^{-3}}$ \\
    $\eta$ & Propagation transition parameter$^{\,c}$ & $0.3$ & $0.3$ \\
    $D_0$ & CR diffusion coefficient normalization$^{\,d}$ & $3\times 10^{28} \;\! \mathrm{cm^{2} \;\! s^{-1}}$ & $3\times 10^{26} \;\! \mathrm{cm^{2} \;\! s^{-1}}$ \\
    $A$ & Boundary flux normalization$^{\,e}$ & $1.0$ & $1.0$ \\
    \bottomrule
  \end{tabular}
  \caption{Fiducial parameters adopted for molecular \textit{clump} and \textit{cloud} models. We define the transport transition by a critical density $n_{\rm H}^{\rm c}=\eta\,n_0$, below which CRs are treated as predominantly diffusive and above which they can stream ballistically (\emph{cf.} boundary-layer/CR-exclusion treatments discussed in the literature$^{\,c}$).
  Quantities not listed are derived from these inputs (e.g. the ballistic-zone radius follows from the mean density, the propagation transition parameter and the cloud’s density profile). 
  \\
  \footnotesize
  \textit{Notes:} \\
  $^{a}$ Distances representative of nearby Perseus/Taurus molecular cloud complexes~\citep[e.g.][]{Yang2023NatAs,Cahlon2024ApJ}. \\
  $^{b}$ Typical radii and densities for molecular clouds and clumps in the Milky Way are informed by standard choices and Galactic samples~\citep{Bergin2007ARAA, Urquhart2014MNRAS, Chen2020MNRAS}. \\
  $^{c}$ The specific value of $\eta$ is a modeling choice used to parametrize a thin transition layer; conceptually similar CR diffusion layers and ``exclusion/penetration'' zones are discussed in, e.g., ~\citealt{Ivlev2018ApJ, Dogiel2018ApJ, Yang2023NatAs, Chernyshov2024PhRvD}.\\
  $^{d}$ In the case of the molecular cloud, the canonical value for the ISM is adopted. For embedded clumps, smaller values due to elevated CR scattering have been proposed, with suppression by around two orders of magnitude reported by e.g.~\citealt{Yang2023NatAs}. \\
  $^{e}$ Introduced as a free parameter later used in fitting analyses to absorb energy-independent normalization effects that do {not} fall within the scope of this study and do not affect our conclusions or interpretation. These include the effects of magnetic mirroring and focusing~\citep{Desch2004ApJ, Owen2021ApJ}, variations in the CR sea level~\citep{Casanova2010PASJ, Aharonian2020PhRvD, Peron2022A&A}, and differences in geometries arising from comparing 2-dimensional plane-of-sky maps with 3-dimensional model configurations. 
  \normalsize}
  \label{tab:fiducial_model}
\end{table*}

\subsection{Methodology}

To calculate the distribution of CR protons and electrons throughout our molecular cloud model, we numerically solve Equations~\ref{eq:transport_equation_protons} and~\ref{eq:transport_equation_electrons}. The boundary condition is 
given by the inward irradiating flux at 
$r = R$. 
The irradiating interstellar CR fluxes for protons and electrons are obtained using \textsc{Galprop} (see Appendix~\ref{sec:appendix} for details), which we adopt as a representative local interstellar spectrum (LIS). We solve the transport equations by reducing them to 
an ODE system in radius and 
using a spherically symmetric computational domain based on the cloud configuration defined in Section~\ref{sec:cr_clouds}, discretized with $N_{\rm r} = 100$ radial grid points and $N_{\rm E} = 200$ energy bins. The solution is obtained using a stiff differential equation solver ({\tt RADAU5}; see~\citealt{hairer2010solving}). 

Since the CR electron distribution is partially determined by the secondary electrons injected via pp interactions of CR protons, we first solve the proton transport equation at each position within 
our computational domain before computing the electron distribution. To obtain an energy-dependent source term 
for secondary CR electrons from pp interactions, $\Gamma_{\rm e}^{\rm sec}(E_{\rm e})$, we use the {\tt AAFrag} tool~\citep{Kachelriess2019CoPhC}. This provides interpolated production spectra derived from the pp event generator {\tt QGSJET-II-04m} above interaction energies of $4$~GeV~\citep{Ostapchenko2011PhRvD, Ostapchenko2013EPJWC, Kachelriess2015ApJ}.
 At lower energies (between particle kinetic energies of $0.28\text{--}4$~GeV), the pp interaction can still occur. As this is outside the energy range where {\tt AAFrag} production spectra are available, we instead adopt analytical cross-sections from~\cite{Kamae2006ApJ}, which are valid near the interaction threshold energy. 

After obtaining the steady-state distributions of CR protons and electrons, we calculate the resulting multi-wavelength emission associated with the CR processes. Pion-decay  
$\gamma$-ray emission is calculated using 
{\tt AAFragpy}~\citep{Kafexhiu2014PRD, Koldobskiy2021PhRvD}, while the X-ray synchrotron component is computed using {\tt Naima}~\citep{Aharonian2010PhRvD, naima}. Having solved the CR electron and proton distributions in our cloud and clump fiducial models, we compute the resulting ionization rates 
throughout the structures by integrating the local CR flux over the relevant ion-producing 
cross-sections. 

CR ionization is driven 
predominantly by the low energy component of the CR spectrum and proceeds through multiple channels, including direct ionization and dissociative ionization of 
H$_2$ by both CR protons and electrons~\citep{Padovani2009A&A}. We include these 
contributions together 
with double ionization with the appropriate weighting (counting two ion pairs per event). 
Additionally, each ionization event can lead to further ionization events, either from knock-on electrons forming an 
additional population of ionizing agents within a cloud, 
or from a single CR interacting multiple times before losing its energy. 
We account for this effect 
by adopting an energy-dependent correction factor $(1+\phi_i)$ for ionization by some species $i$, which scales the cross-sections for multiple ionization channels. 

The overall ionization rate caused by CRs of species $i$ is then given by 
\begin{align}
\zeta(r)
&= \sum_{i = \mathrm{p},\mathrm{e}}
\int dE_{i} \;
J_i\!\left(E_{i},{r}\right)\,
\left[1+\phi_i(E_{i})\right]\,
\sigma_{i,\mathrm{ion}}(E_{i})\ ,
\label{eq:zeta_def}
\end{align}
where $J_i(E_{i}, r)$ is the flux of CRs of species $i$ of energy $E_{i}$ at position $r$ within the cloud. In our treatment, charge-exchange contributions are not considered. We calculate this function for both CR electrons and protons separately for our two fiducial (cloud and clump) models. The total ionization rate is then reported as the sum of these two contributions. The multiplicity function $\phi_i$ has been quantified up to $~\sim 100$ MeV, and becomes very weakly energy-dependent at higher energies. We approximate it for protons by using a linear piecewise function, 
with a lower cut-off of 0.1 MeV, as indicated by the results of~\cite{Glassgold1973ApJ}. Values reported by ~\cite{Cravens1978ApJ} were then used to construct the piecewise function 
up to a maximum energy of 100~MeV, above which their empirical maximum of $\phi_\mathrm{p} = 0.74$ is set as a fixed value with increasing energy. To obtain a corresponding $\phi_\mathrm{e}$ for electrons, we follow ~\cite{Padovani2009A&A} and scale it from the proton multiplicity function as $\phi_\mathrm{e} (E_\mathrm{e}) \approx \phi_\mathrm{p}(E_\mathrm{p} = m_\mathrm{p} E_\mathrm{e}/m_\mathrm{e})$. This is valid in the Bethe-Born approximation, which 
\cite{Cravens1978ApJ} confirm for protons and electrons in the range 1--100~MeV. {Note that, unless otherwise stated, we report $\zeta$ as an ionization rate per H nucleus. Literature values defined per ${\rm H}_2$ 
 molecule are converted consistently when compared to our results.}

\subsection{Fiducial parameters}
\label{sec:fiducial}

As a baseline, we adopt representative models 
for a molecular cloud and an embedded dense clump.  
All fiducial parameters are listed in Table~\ref{tab:fiducial_model}.  
We set the cloud distance to 150~pc. This is comparable to some of the nearest star-forming molecular cloud complexes to Earth, such as Taurus~\citep{Cahlon2024ApJ}.  
Cloud and clump radii and densities are representative of the 
hierarchical components of the ISM, informed by standard descriptions \citep[e.g.][]{Bergin2007ARAA} and are consistent with the properties of clouds reported by surveys of nearby giant molecular clouds and clumps \citep{Urquhart2014MNRAS, Chen2020MNRAS}. 

For molecular clouds, 
we adopt the canonical Galactic ISM value for the diffusion coefficient normalization in Equation~\ref{eq:diff_coeff}.  
Constraints from CR secondaries (e.g. B/C) imply $D_0 \sim$ a few $\times 10^{28}~\mathrm{cm^2\,s^{-1}}$ at energies of a few GeV \citep[see][for reviews]{Berezinskii1990book, Strong2007ARNPS}.  
Observations of nearby molecular clouds and their embedded clumps suggest that the value of $D_0$ may be strongly suppressed in dense gas, in some cases by factors of $\sim 100$ \citep[e.g.][]{Yang2023NatAs}.  
We therefore adopt a reduced $D_0$ inside the embedded clump model accordingly as part of our fiducial setup.

We consider three propagation scenarios:  
(i) a fully ballistic model,   
(ii) a fully diffusive model, and   
(iii) a two-zone model where transport transitions from diffusion-dominated to quasi-ballistic, following the schematic in Figure~\ref{fig:cloud_framework}.  
In the two-zone case, CRs enter from the ambient Galactic sea into an outer diffusive layer, and then traverse an inner region where the transport is ballistic.  
We parametrize the location of this transition by $\eta$.  
This boundary is a modeling choice. It is not intended to represent a sharp physical discontinuity.  
Rather, it provides a simple 
way to capture the net effect of a CR ``penetration/exclusion'' layer discussed in the literature \citep[e.g.][]{Ivlev2018ApJ, Dogiel2018ApJ, Yang2023NatAs, Chernyshov2024PhRvD, Axen2024ApJ}, 
which may be an 
ion–neutral damping threshold, wave growth vs damping threshold, or may be related to ionization fraction. We  assess the impact of its presence on the resulting non-thermal emission signatures from CRs in a molecular cloud.

We also include a boundary flux normalization factor, $A$.  
In our calculations, $A$ does not have a unique physical interpretation.  
It absorbs energy-independent normalization effects that we do not model explicitly and that are not central to the conclusions presented in this work.  
Later, we treat $A$ as a free parameter in fitting to account for such effects without physically modeling them.  
These include magnetic mirroring and focusing \citep{Desch2004ApJ, Owen2021ApJ}, spatial variations in the CR sea level \citep{Casanova2010PASJ, Aharonian2020PhRvD, Peron2022A&A}, and geometric differences between projected maps and our idealized 3-dimensional model configuration, all of which can modify the interpreted normalization of the CR flux irradiating a molecular cloud structure.

\section{Results}
\label{sec:results}  

\subsection{$\gamma$-ray emission}
\label{sec:gammarays}

\begin{figure}
    \centering
    \includegraphics[width=\linewidth]{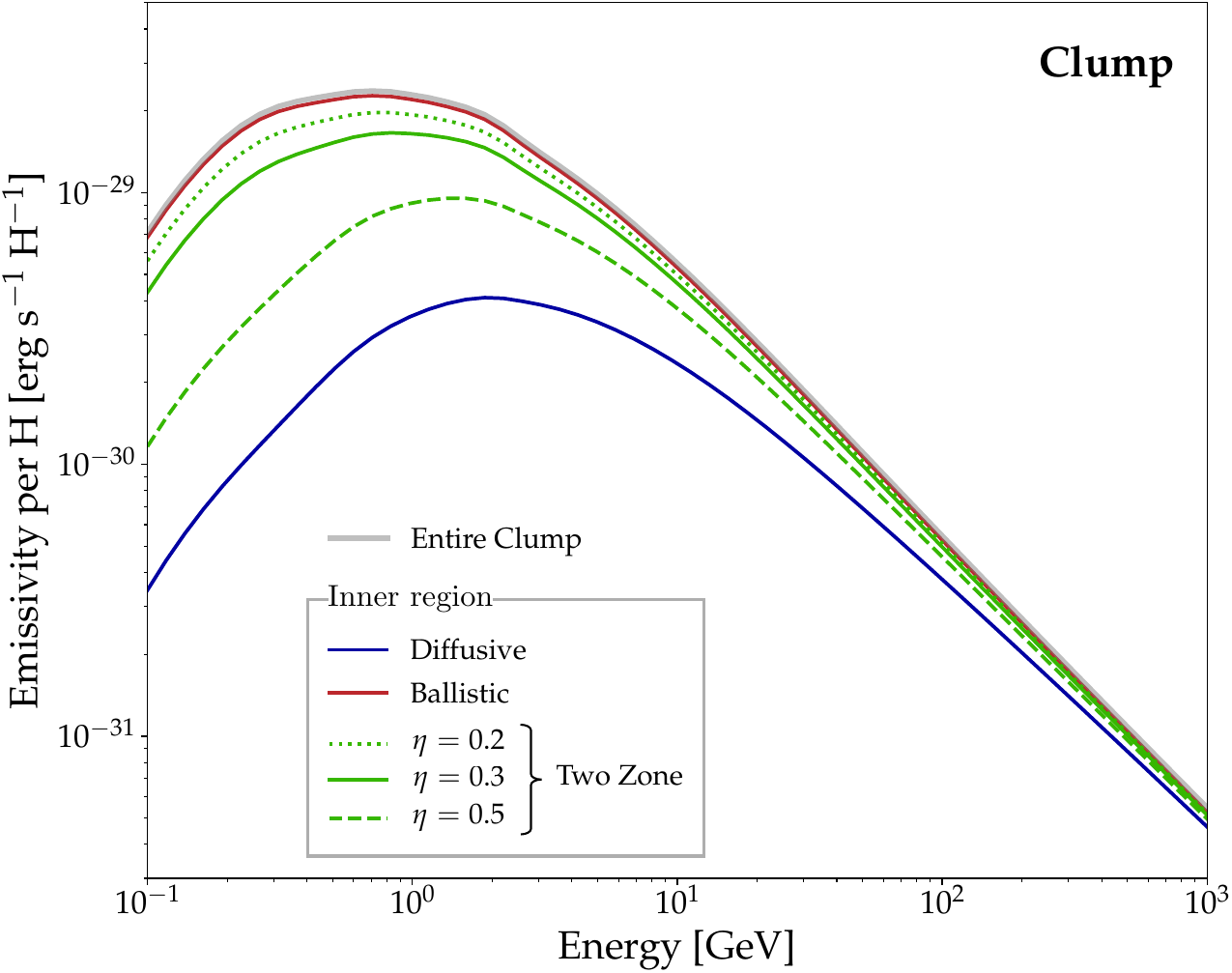}
    \caption{Pion-decay $\gamma$-ray emissivity per H atom produced by CR protons in the fiducial clump model. The thick gray line shows the emissivity integrated over the full clump. Other lines show the contribution from the inner $r<0.25\,\mathrm{pc}$ region, comparable to the smallest angular scale that could be resolved for a clump at $d=150\,\mathrm{pc}$ in this energy range with an instrument like \textit{Fermi}--LAT. {Colored curves indicate the CR transport model: fully diffusive (blue), ballistic (red), and two-zone models (green) with transition parameter $\eta$ (values as labeled).} 
    In the ballistic case, the inner-region spectrum closely tracks the clump-integrated emissivity. As the extent of the diffusive layer increases (larger $\eta$ to the fully diffusive limit), the core emission below $\sim 100\,\mathrm{GeV}$ becomes progressively suppressed. This effect arises due to the enhanced magnetic scattering the CR protons experience in the diffusive layer, which boosts their attenuation by the pp interaction. Losses are more severe at lower energies where CRs diffuse more slowly.}
    \label{fig:gamma_spec_clump} 
\end{figure}

\begin{figure}
    \centering
    \includegraphics[width=\linewidth]{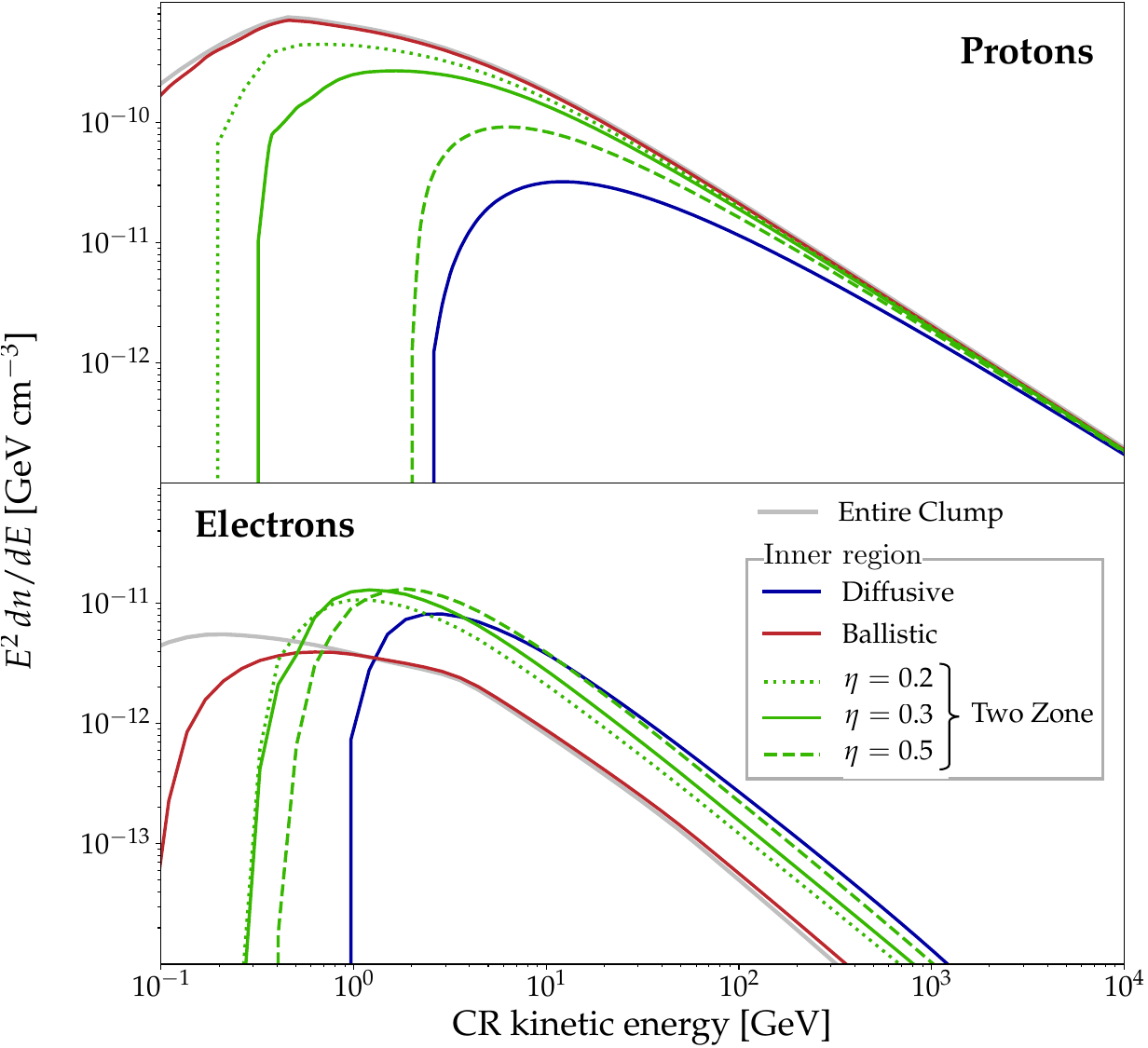}
    \caption{{Volume-averaged CR proton (upper panel) and electron (lower panel) spectra within the inner $r<0.25\,\mathrm{pc}$ region for each CR transport scenario. The grey curve indicates the interstellar boundary spectrum ({\tt GALPROP} LIS). The diffusive case (blue) shows strong low-energy suppression due to energy-dependent exclusion, while the ballistic case (red) preserves the boundary spectrum with minimal modification. The two-zone models (green) produce intermediate levels of suppression depending on the propagation transition parameter, $\eta$. In the electron panel, the clump-interior spectra can exceed the ISM boundary at intermediate energies due to in-situ production of secondary electrons from CR proton interactions with the dense gas.}}
    \label{fig:cr_spec_clump_elec_prot}
\end{figure}

We compute the $\gamma$-ray emission associated with the steady-state CR proton distribution for our fiducial cloud and clump
models. We find that the resulting emissivity
for our fiducial cloud model shows little dependence
on the transport configuration, due to weak hadronic
attenuation in the diffuse cloud gas. However, 
the effects of CR transport variation are much more apparent on the clump scale. 
Our results are shown in 
Figure~\ref{fig:gamma_spec_clump}, which demonstrates the 
effect of strong CR 
scattering in a diffusion layer 
on the $\gamma$-ray emission from a clump, where the pion-decay emissivity per H atom from the integrated 
spectrum with the contribution from the central $r<0.25\,\mathrm{pc}$ region of a clump (with our fiducial parameters). We adopt this inner radius as a pragmatic baseline observational reference scale. For a clump at a distance of 150 pc, 
it corresponds to an angular size that is comparable to the smallest spatial scale that \textit{Fermi}--LAT could plausibly discriminate in this energy range. 
This comparison therefore provides a direct way to gauge whether energy-dependent transport effects are expected to imprint measurable spectral differences between the clump core and the clump as a whole, rather than being fully washed out by angular averaging. For a 
typical clump configuration, we 
expect that the $\gamma$-ray emission volume to pion decays 
would be 
dominated by the outer regions of the clump structure. 
These would be less affected by spectral differences arising due to differences in CR transport, which would 
emerge predominantly in the central core regions where CRs at lower energies can be strongly excluded (see also this 
pattern in the radial profiles shown in Figures~\ref{fig:diff_suppression_ionization} and~\ref{fig:change_eta}). 
{The corresponding volume-averaged CR spectra in the same inner region are shown in Figure~\ref{fig:cr_spec_clump_elec_prot}. 
These spectra show that the low-energy $\gamma$-ray suppression arises from energy-dependent filtering of the parent CR proton population in the diffusive transport cases, and 
that this effect becomes more severe as the extent of the diffusive transport layer increases. 
They also show that secondary CR electrons are enhanced relative to the incident interstellar spectrum at intermediate energies when CR proton attenuation is more severe (see also Section~\ref{sec:xrays}).}

Our results show that spectral suppression emerges when severe magnetic scattering of CR protons is operating. In the ballistic limit, the inner-region emissivity closely tracks the clump-integrated spectrum, indicating that CR protons are able to 
efficiently penetrate into the center with little energy-dependent filtering. 
By contrast, the 
two-zone and fully diffusive models exhibit a progressively stronger suppression of the core emissivity below  
100 GeV 
as the volume-filling extent of the diffusive layer increases (larger $\eta$ and, ultimately, the fully diffusive case).  
Physically, enhanced scattering in the diffusive layer increases the effective residence time and path length of CR protons in dense gas, raising the probability of pp interactions before the particles can reach the core.  The effect is strongest at lower energies, where diffusion is slower, so attenuation in the surrounding material more efficiently depletes the low-energy proton population 
that would otherwise 
be available to power the inner region's pion decay emission. 
The corresponding calculation for the larger fiducial cloud (not shown) indicates negligible suppression would be expected for any transport model, consistent with its lower attenuating column and the more rapid CR propagation expected through the lower-density medium.

The presence of a CR exclusion zone 
when substantial scattering in an extended diffusion zone arises is 
consistent with the sub-10-GeV shielding and low-energy $\gamma$-ray spectral suppression reported by earlier studies in regions with high attenuation columns~\citep[e.g.][]{Yang2023NatAs, 2024A&AKamalYoussef, Zeng2024RAA, Jiang2025MNRAS}. Consistent with these earlier studies, our results confirm that 
such effects are only able to arise when CR diffusion operates at a level where charged particle scattering in a diffusion region of a cloud is operating at several orders of magnitude higher than that typical of the Galactic ISM overall. One interpretation of this is that magnetic turbulence may persist in dense interstellar gas structures, despite ion-neutral damping. 

This presents an intriguing alternative interpretation of the dearth of detected $\gamma$-ray emission from diffuse, multiphase extra-galactic environments such as the circum-galactic medium and intra-cluster medium. In general, hadronic $\gamma$-ray emission tracks both the CR density and the gas density in a given region. However, if CR transport is sufficiently rapid that CRs decouple from, and escape from, cold dense structures, the mass-weighted emissivity can be reduced even when a substantial gas reservoir is present~\citep{Habegger2025arXiv251024622H}. In this picture, dense phases may remain $\gamma$-ray faint because the local CR energy density is depressed relative to the ambient medium. 
Our results highlight a complementary possibility in the opposite transport limit. If CR coupling and scattering in and around dense structures is particularly strong, CR protons can be attenuated in surrounding diffusive envelopes such that they do not efficiently penetrate to the densest clumps. This ``shielding'' 
selectively suppresses the contribution 
of the highest-density phase and can therefore reduce the region's mass-weighted $\gamma$-ray emissivity. The degree to which the total $\gamma$-ray luminosity is then also suppressed would depend on 
the fraction of the gas mass residing in the shielded phase and on the effective grammage accumulated in the enveloping medium, which increases when diffusion is slow. In the extreme case, where a large fraction of the mass is locked into compact, very dense clumps, the densest gas could remain effectively $\gamma$-ray ``hidden'' to hadronic probes~\citep[e.g.][]{Abrahams2017ApJ, Lai2026PhRvD}. These two scenarios are, in principle, distinguishable: strong envelope shielding generically imprints a pronounced GeV-band energy-dependent suppression, whereas rapid escape and decoupling more closely resembles a low-grammage reduction with a weaker and less distinctive spectral signature.

\subsection{X-ray emission}
\label{sec:xrays}

\begin{figure}
    \centering
    \includegraphics[width=\linewidth]{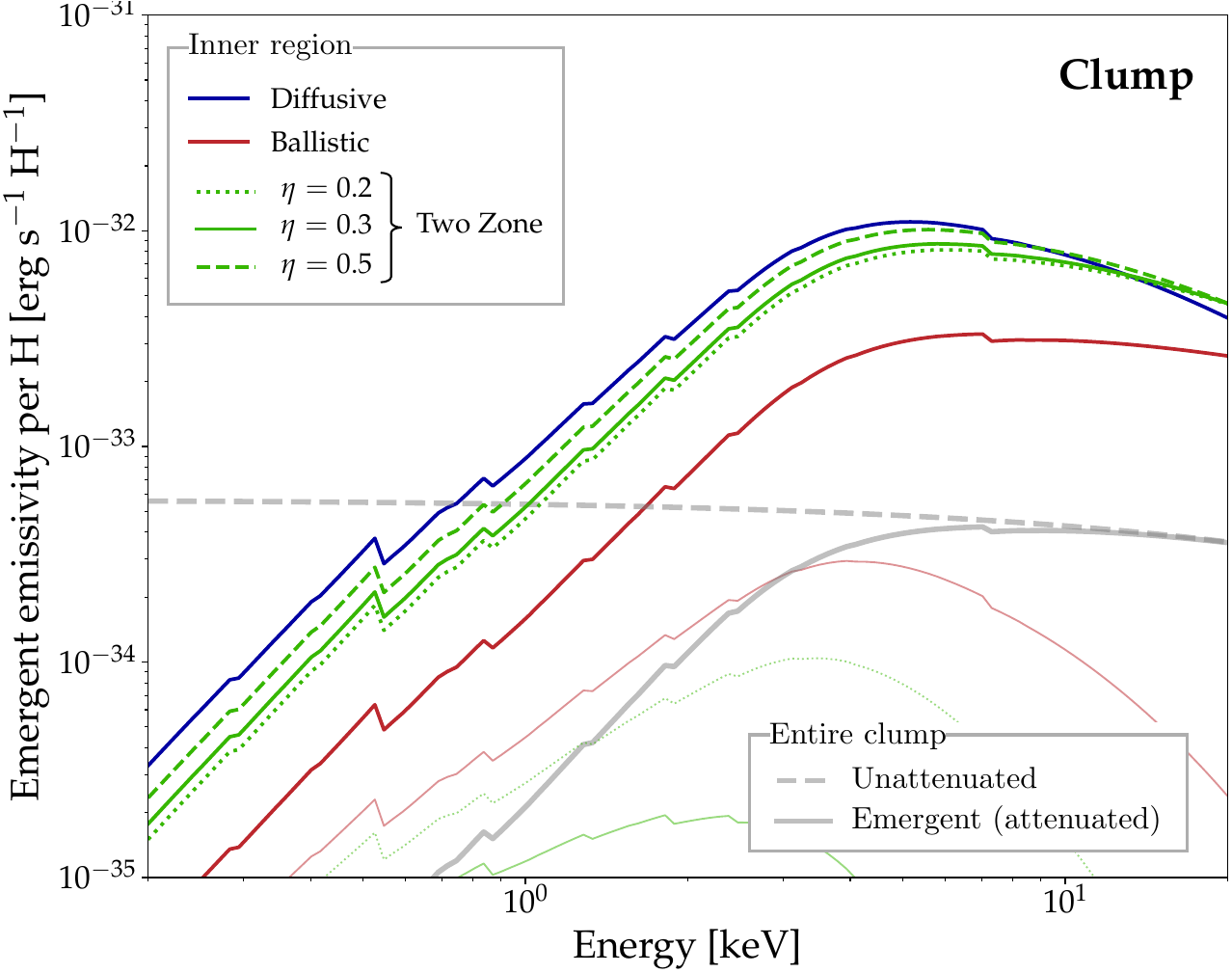}
    \caption{Emergent synchrotron X-ray emissivity per H atom from {primary and secondary electrons} produced in hadronic interactions in the fiducial clump model. Colored curves show the contribution from the inner region ($r<0.25\, \mathrm{pc}$) for different CR transport prescriptions: fully diffusive (blue), ballistic (red), and two-zone models (green) with transition parameter $\eta$ (values as labeled). Photoelectric absorption within the clump is applied, so the colored curves represent the emergent (attenuated) emissivity. The gray curve shows the clump-integrated spectrum for reference: intrinsic (unabsorbed; dashed) and emergent (attenuated; solid). {Thick lines show the total emission from primary and secondary electrons. Thin faint lines show the sub-dominant contribution from primary electrons only, with the difference due to the emission from secondary electrons. We find that the secondary-driven component is dominated by  electrons of energies $\sim$ 0.1-1 TeV. }}
     \label{fig:xray_spec_clump} 
\end{figure}

We compute the synchrotron emission associated with the steady-state CR electron distribution for our fiducial cloud and clump models.  We put focus on the hard X-ray regime, as soft X-rays are strongly suppressed by photoelectric absorption along typical Galactic lines of sight. As with the $\gamma$-ray emission, we find that the resulting X-ray emissivity for our fiducial cloud model is low and shows little dependence on the transport configuration, consistent with the weak hadronic attenuation inferred from the corresponding $\gamma$-ray results. In contrast, for the molecular clump case, the presence of any diffusive zone produces a pronounced enhancement of the emergent hard X-ray emissivity relative to the purely ballistic model. After absorption is applied, the spectra remain strongly suppressed at soft energies, but the separation between ballistic and non-ballistic transport remains clear above $\sim$keV energies. This is presented in Figure~\ref{fig:xray_spec_clump}.  
For consistency with our $\gamma$-ray calculations, we report the emission as an emissivity per Hydrogen atom and consider both (i) emission integrated over the whole clump, and (ii) the ``inner-region'' aperture intended to mimic a resolved core measurement, with the same 0.25 pc extent as considered in our $\gamma$-ray calculations in Section~\ref{sec:gammarays}. {For comparison, we also show the contribution from primary electrons, which is sub-dominant compared to the emission driven by secondary electrons.}

\begin{figure*}
    \centering
    \includegraphics[width=0.48\linewidth]{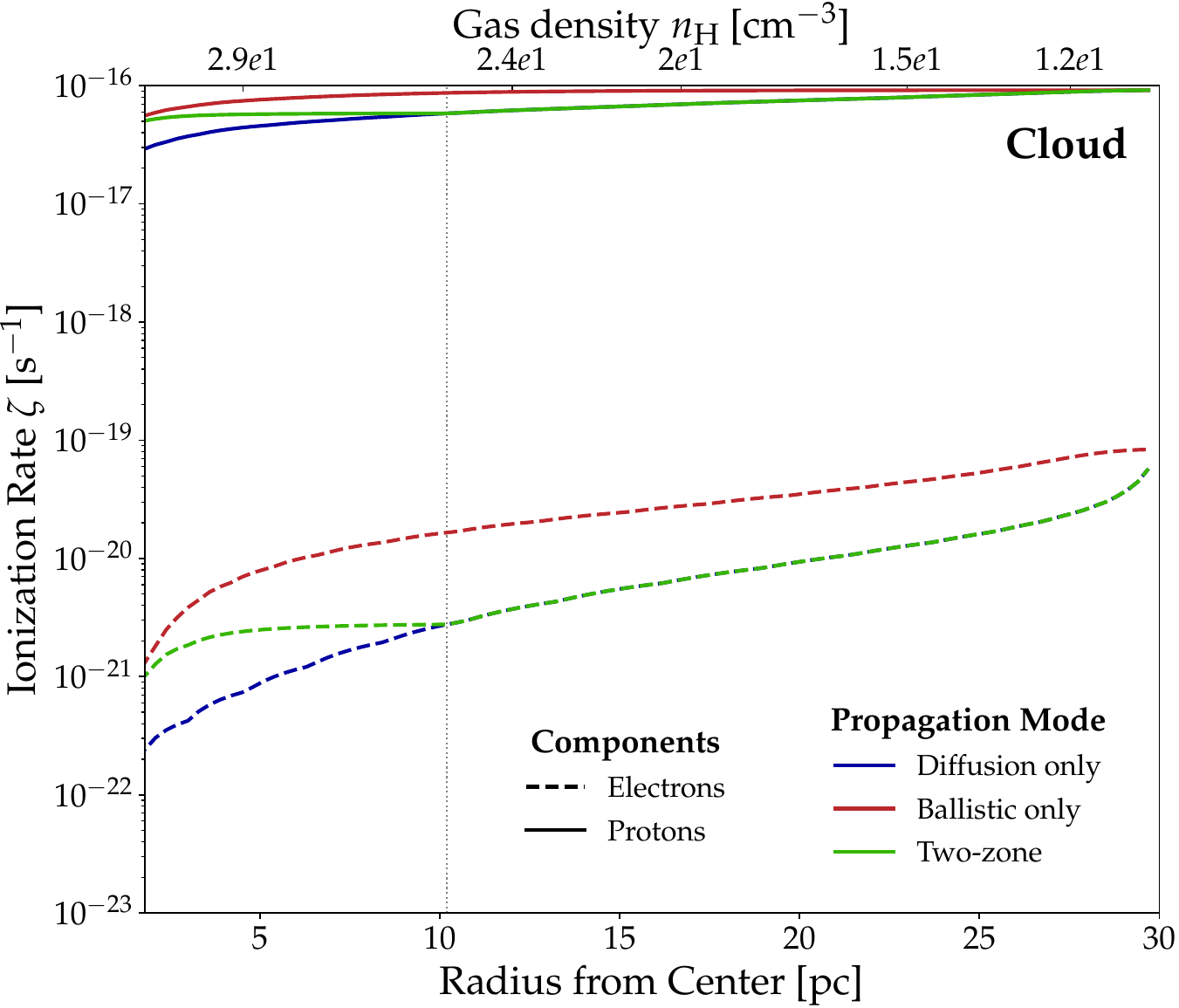}
    \hspace{0.2cm} \includegraphics[width=0.48\linewidth]{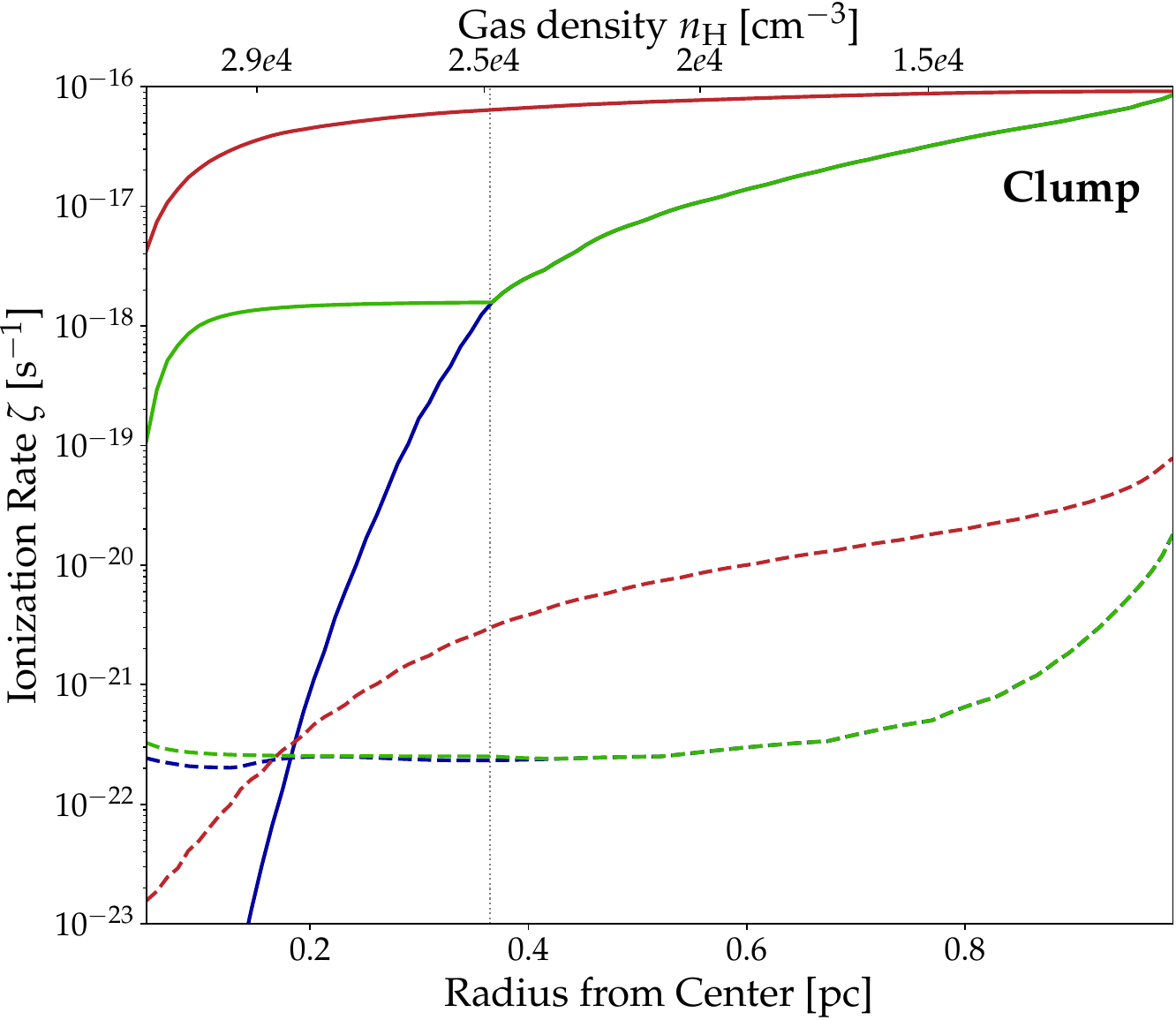} 
    \caption{CR ionization rate profiles, $\zeta(r)$, for the fiducial molecular cloud (left) and a clump (right). In both cases, for comparative purposes, the LIS is set as the irradiating external CR flux. 
We compute $\zeta$ by integrating the local CR proton and electron fluxes over ion-producing cross-sections
(direct and dissociative ionization of H$_2$, and weighted double ionization), including secondary ionizations.
Curves compare three transport prescriptions: fully diffusive (blue), fully ballistic (red), and a two-zone model (green)
with an outer diffusion-dominated layer and an inner quasi-ballistic region; the adopted transition parameter is $\eta=0.3$
(cf.\ Figure~\ref{fig:cloud_framework}), with the position of the transition indicated by the vertical dotted gray line. The upper axis shows the gas number density at a given position, according to the adopted profile $n_{\rm H}(r)$. 
Line styles separate proton (solid) and electron (dotted) contributions, including secondary electrons from hadronic interactions. 
For the fiducial parameters, the total ionization rate is proton-dominated. It varies only weakly with depth in the low-density cloud, with values close to the canonical \textit{Spitzer} value. In clumps, the ionization rate shows appreciable strong variation with the underlying CR transport properties.}
    \label{fig:fiducial_ionization_results} 
\end{figure*}

To connect the intrinsic synchrotron spectrum to an emergent quantity, we fold in photoelectric absorption using an energy-dependent transmission computed with a TBabs-like cross-section, $\sigma_{\rm pe}(E)$ with Wilms abundances \citep{Wilms2000ApJ}. The attenuation is then applied internally using a ``slab'' prescription appropriate for emission distributed throughout the absorber, $T_{\rm int}(E)=\left[1-\exp \left(-\tau_{\rm emit}\right)\right]/\tau_{\rm emit}$ with $\tau_{\rm emit}(E)=N_{\rm H,emit} \sigma_{\rm pe}(E)$. In this emissivity-per-H representation we do not apply any additional foreground screen external to the cloud (i.e. the effective foreground column is set to zero), so the plotted spectra isolate attenuation by the model cloud itself. For reference, we also show the intrinsic (unabsorbed) emissivity for the full cloud, allowing transport-driven spectral changes to be visually separated from absorption effects.

Our results show that substantial CR scattering in a diffusive envelope 
boosts the pp interaction rate in that region. In addition to suppressing the pion-decay $\gamma$-ray emission at low energies, this effect increases secondary CR electron injection into the most magnetized regions of the clump structure, driving a noticeable increase in hard X-ray emission. In models where CR transport is at least partly diffusive, CRs accumulate a larger effective grammage as they traverse the cloud envelope. This increases the hadronic interaction rate in the envelope relative to the purely ballistic limit, which increases the injection of secondary electrons produced in charged-pion decays. 
{This effect can be seen more directly in the CR spectra shown previously, in Figure~\ref{fig:cr_spec_clump_elec_prot}. 
Because secondary electrons} are deposited preferentially where the gas density and cloud magnetic field strengths are highest, their radiative losses are dominated by synchrotron cooling. The highest-energy secondaries can therefore contribute a non-thermal synchrotron component extending into the X-ray band. In our two-zone results (green lines in Figure~\ref{fig:xray_spec_clump}), our calculations show that varying the transition parameter $\eta$ (which controls the extent of the diffusive envelope in the two-zone model) produces a comparatively modest change in the emergent hard X-ray spectrum, compared with the much stronger ballistic-versus-diffusive contrast. In this sense, synchrotron X-rays provide a cleaner diagnostic of whether a clump hosts a diffusion-dominated envelope at all, while the detailed value of $\eta$ is more efficiently constrained by the shape and energy dependence of the $\gamma$-ray suppression.

Our results suggest that a detection of hard X-ray emission from a nearby, otherwise passive clump would constitute compelling evidence for non-ballistic CR transport and strong coupling of CRs with slow diffusion operating in the clump envelope. Conversely, stringent upper limits would firmly disfavor extended diffusion zones, and favor a scenario where ballistic transport operates throughout a large volume fraction of the cloud, especially if robust magnetic field measurements for the particular structure were available. $\gamma$-ray and ionization-based constraints on CR transport in dense ISM substructures would provide complementary information about the prevalence and spatial extent of a diffusion zone if one were identified from excess X-ray synchrotron emission. Although X-ray emission has not yet been reported in molecular clouds or their substructures, even in the vicinity of supernova remnants where CR irradiation may be higher~\citep{Aharonian2013APh, Tsuji2024ApJ}, the prospects to detect such a signature in nearby molecular clouds and clumps with future instruments are shown in Section~\ref{sec:illustrative_applications}. 

\subsection{CR ionization profiles}
\label{sec:ionization_profiles}

In Figure~\ref{fig:fiducial_ionization_results}, we show the CR ionization rate profile,  
$\zeta(r)$ for our fiducial cloud and clump models (Table~\ref{tab:fiducial_model}), integrated over CR proton and electron energies of $E=0.3~{\rm keV}$--$1~{\rm PeV}$. This 
captures the highest energy CRs of interest in this work, 
and extends down to the lowest energy CRs that could have an impact on our results. At energies below 0.3 keV, the CR flux is small and highly dependent on the exact configuration of the individual cloud, with their effects not having a discernible impact on the total ionization rate. 
In both cases, the LIS is adopted as the irradiating CR flux. While this is reasonable in the case of isolated clumps and clouds within the ISM, 
the hierarchical configuration of interstellar clouds 
would suggest that clumps are often found embedded within clouds. Because of this, a strict treatment would 
consider the attenuating and reprocessing effect the 
diffuse cloud environment would have in modifying the LIS before it reached the external boundary to irradiate a clump. Such complexities are not explicitly treated in this work, 
although the effects of parent cloud attenuation is later parametrised when applying our model to the Taurus molecular cloud complex (see Section~\ref{sec:illustrative_applications}). 
For our fiducial model, our results are intended to demonstrate the effects of CR propagation on possible  
observables, so we do not put focus on the distortion of the LIS as it penetrates a cloud to irradiate a clump.

For our parameter choices,
the total ionization rate is dominated by CR protons in both structures, with electrons remaining sub-dominant at all radii in the cloud, and only exceed the effect of CR protons in dense clumps when the degree of CR scattering (via suppression of the CR diffusion coefficient) is very severe. The total ionization rate profiles vary  
appreciably between the {clump and cloud structures}. It decreases only  weakly with depth in the low-density cloud, with values close to the canonical Milky Way value of $\sim 10^{-17} \;\! {\rm s^{-1}}$ {per H} (the \textit{Spitzer} value; e.g.~\citealt{Glassgold1973ApJ,Cummings2016ApJ}). In clumps, the ionization rate shows strong suppression in the central, densest regions, with the severity of the suppression being closely dependent on the underlying CR transport properties. This is, in part, due to strong ionization cooling the CRs experience as they traverse large column densities into the densest regions. The effective column they encounter increases with the severity of magnetic scattering they undergo in the outer diffusive transport layer. The slower transport invoked in our 
clump 
model compared to the cloud model also 
allows additional electron ionization to develop. This is because the higher-energy population of 
CRs in the outer diffuse envelope undergoes more hadronic interactions and boost secondary electron injection. This associated secondary electron population may provide a further, independent probe of CR transport on clump scales at high energies (see Section~\ref{sec:x_rays}). 

\begin{figure}[t!]
    \centering
    \includegraphics[width=\linewidth]{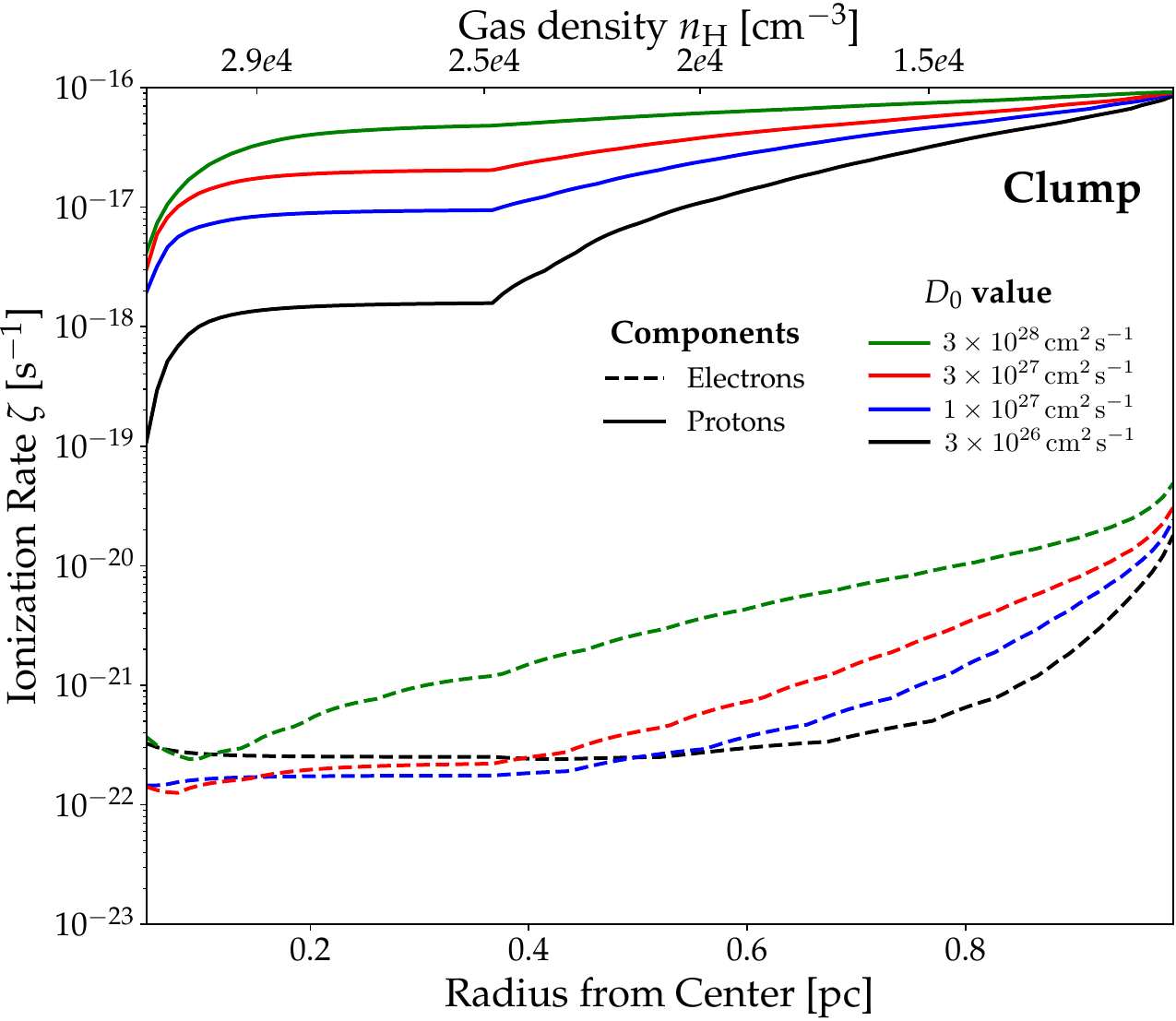}
    \caption{Ionization-rate profiles, $\zeta(r)$, through the fiducial clump model, showing the impact of suppressing the diffusion coefficient normalization $D_0$ in the diffusive transport layer due to varying CR scattering rates in magnetic turbulence. Curves show the proton (solid) and electron (dotted; including secondaries from hadronic interactions) contributions to $\zeta$ as a function of radius (lower axis), with 
    the corresponding gas density $n_{\rm H}(r)$ shown on the top abscissa. Colors indicate 
    $D_0 = 3\times 10^{28}$ (green), $3\times 10^{27}$ (red) $1\times 10^{27}$ (blue), and $3\times 10^{26} \, \mathrm{cm}^2 \mathrm{s}^{-1}$ (black) diffusion coefficients, corresponding to the ISM canonical choice and suppression factors of 10, 30 and 100, respectively. Increasing suppression of $D_0$ yields stronger attenuation of the ionizing CR flux at a given depth and hence a lower $\zeta$. Very strong suppression associated with low diffusion coefficients, coupled with secondary electron-driven ionization, begins to change the balance of CR species driving the ionization process. The transition between the diffusion and ballistic transport zones is set at our fiducial choice of $\eta = 0.3$ in all cases.\\
    Alt text: Cosmic ray ionization profiles in a clump with different levels of suppression of the diffusion coefficient.}
    \label{fig:diff_suppression_ionization}
\end{figure}

\begin{figure}[t!]
    \centering
    \includegraphics[width=\linewidth]{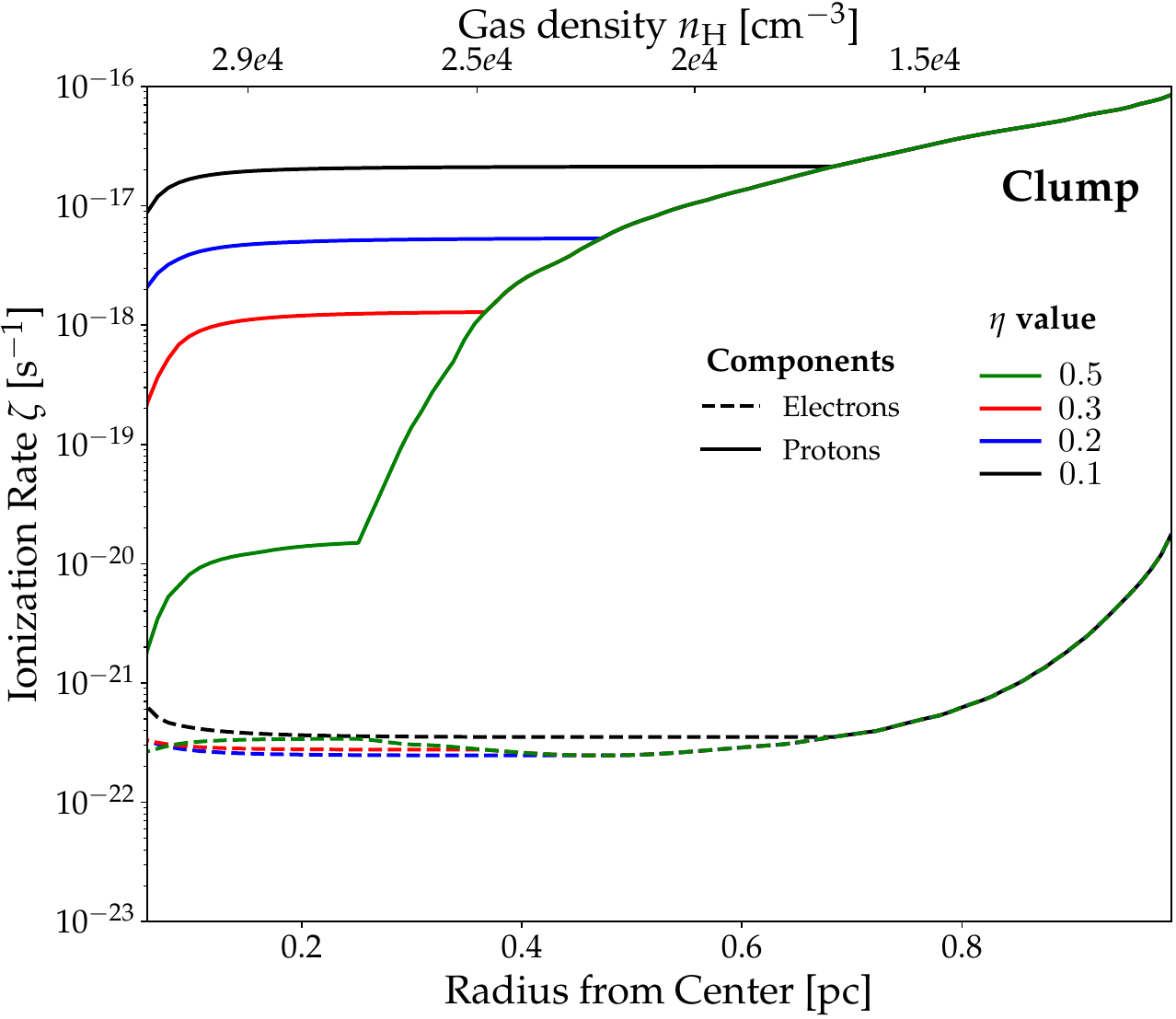}
    \caption{As Figure~\ref{fig:diff_suppression_ionization}, but varying the two-zone transition parameter $\eta$, 
    at which CR transport switches from diffusion-dominated in the outer layer to quasi-ballistic propagation in the inner region. Curves show the proton (solid) and electron (dotted) contributions for $\eta = 0.1$ (black), 0.2 (blue) 0.3 (red; fiducial choice), and 0.5 (green). The diffusion normalization is held fixed at the fiducial clump value of $D_0 = 3\times 10^{26} \, \mathrm{cm}^2 \mathrm{s}^{-1}$ (Table~\ref{tab:fiducial_model}). Larger values of $\eta$ (representing a more extended diffusion zone) leads to more scattering of CRs and stronger cooling and attenuation of the ionizing CR population into the core. This leads to a lower central value of $\zeta$.}
    \label{fig:change_eta} 
\end{figure}

The impacts of slow diffusion on ionization are explored explicitly in Figure~\ref{fig:diff_suppression_ionization}, where 
the electron and proton ionization profiles for different suppression factors of the CR diffusion coefficient are presented. In all cases considered, it can be seen that more suppression of the diffusion coefficient 
results in more attenuation of the CR flux at a given 
density or position, and a suppressed ionization rate at that location. At very strong suppression factors (for choices of $D_0$ below $\sim 3 \times 10^{26} \, \mathrm{cm}^{2} \mathrm{s}^{-1}$; see the black line in Figure~\ref{fig:diff_suppression_ionization}), there is a noticeable increase in the 
prevalence of secondary CR electron-driven ionization, 
which could eventually dominate in these strong scattering regimes at higher densities, 
and would become more abundant as the CR scattering rate is increased further (i.e. if $D_0$ is suppressed more). 

Expanding the diffusion zone where CRs 
undergo strong scattering in magnetic turbulence 
similarly operates to boost the interaction rates and cooling effect the CRs experience as they 
propagate into a cloud. This is shown in Figure~\ref{fig:change_eta}, where we vary the transition parameter $\eta$ that sets the density threshold $\eta n_0$ of the transition from diffusion in the outer envelope to quasi-ballistic propagation in the inner region. As $\eta$ increases, the diffusion-dominated layer extends to higher densities (equivalently, the ballistic core shrinks). CRs therefore accumulate a larger effective grammage before reaching the densest gas, which amplifies their losses and suppresses the contribution from low-energy primary CRs at small radii. At the same time, the larger diffusive volume increases the probability of hadronic interactions for the high-energy proton population, injecting a larger secondary electron component. This secondary population can partially offset the attenuation of primaries and can become increasingly important in the highest-density regions, producing a flatter ``floor'' in the total ionization rate and shifting the balance from proton-dominated ionization in the outer clump toward electron-dominated ionization deeper in the core.

This behaviour highlights that, 
 even at fixed external CR boundary conditions, the ionization state of dense substructures can be regulated strongly by the amount of scattering and, hence, attenuation experienced by low-energy CRs before they can enter and traverse the inner regions of a cloud. In our clump model, increasing the level of magnetic scattering either by suppressing the diffusion coefficient (Figure~\ref{fig:diff_suppression_ionization}) or by extending the outer diffusion-dominated layer to higher densities via larger $\eta$ (Figure~\ref{fig:change_eta}) increases the effective grammage accumulated prior to reaching the densest gas. 
 This preferentially excludes and cools the low-energy primary CR population, producing a steep suppression of the proton-driven ionization rate toward the core in the strongly diffusive limit, and can strongly reshape 
the CR spectrum and ionization profile inside dense gas, including changing which species dominates the ionization in the core. 
 These effects 
 are qualitatively expected, given 
the anti-correlation reported between ionization rates and gas column densities~\cite[see][for an overview]{Albertsson2018ApJ}. However 
 this finding 
 is also supported by recent  detailed  
observational studies that map the variation of the CR ionization rate through specific molecular cloud structures. These have revealed noticeable variation of the ionization rate within individual dense ISM clouds, for example in NGC 1333~\citep{Pineda2024A&A}, Orion OMC-2 and OMC-3~\citep{Socci2024A&A}, and some massive pre-stellar clumps~\citep{Sabatini2023ApJ}. 
As these are presumably subject to the same irradiating CR spectrum, 
such variations can be more naturally attributed to variations in CR transport and attenuation 
throughout these clouds' complex structure, instead of 
alternative possibilities that could account for variations of $\zeta$ as measured 
more generally across the Galactic ISM, such as in the proximity of nearby particle acceleration sites~\citep[e.g.][]{Phan2023PhRvD, Indriolo2023ApJ, Indriolo2025arXiv}, 
variations in the low energy CR spectrum~\citep[e.g.][]{Padovani2009A&A, Nath2012MNRAS}, interpretations of the underlying astrochemical tracers of $\zeta$ in varying physical conditions~\citep[e.g.][]{Shingledecker2016ApJ, Redaelli2024A&A, Luo2024A&A}, 
or even methodological differences in their interpretation~\citep[see discussion in][]{Obolentseva2024ApJ}. 

\subsection{Illustrative application}
\label{sec:illustrative_applications}

Although our fiducial models highlight the qualitative changes in emission expected under different CR-transport scenarios in molecular clouds and their substructures, it is also important to assess whether these differences are realistically detectable and therefore usable as empirical tests of CR propagation. We therefore present an illustrative application to a well-studied nearby target for which (i) an angular resolution of order $\sim 0.1^{\circ}$ corresponds to physical scales of $\sim 0.2$--$0.5$ pc (i.e. comparable to clump scales; see Table~\ref{tab:fit_results}), and (ii) extensive ancillary data already provide strong spatial templates for the gas distribution and cloud geometry. 
We adopt the Taurus molecular cloud as a representative laboratory. 

The Taurus molecular cloud {is interesting as a laboratory for star-formation and CR transport studies}. At $\sim$ 147 pc~\citep{Zucker2019ApJ}, it is one of the nearest molecular clouds to the Sun. It is relatively quiescent compared to other nearby clouds (dominated by low-mass star formation; for a review, see~\citealt{Kenyon2008hsf1}), while still benefiting from a high-resolution view of its 
 gas and dust column density and 3D structure~\citep{Soler2023A&A}, and well-measured magnetic-field geometry over a broad range of scales~\citep[e.g.][]{Eswaraiah2021ApJ, Doi2021ApJ}. 
The Taurus cloud has been studied in $\gamma$-rays with \textit{Fermi}-LAT at GeV energies~\citep{Neronov2017A&A}, including down to the clump-scale, where $\gamma$-ray substructures have been identified in each system~\citep{Yang2023NatAs}. As such, this cloud is a compelling nearby laboratory for CR microphysics for which ample data is already available, and will be ideal for future detailed multi-wavelength tests of CR transport using the framework set out in this work. 

\subsubsection{Application to the Taurus molecular cloud}
\label{sec:clouds_application}

\begin{table*}
\centering
\small
\setlength{\tabcolsep}{6pt}
\begin{tabular}{l c c c c c c c c c c c}
& \multicolumn{4}{c}{\textit{Adopted parameters}} 
& \multicolumn{7}{c}{\textit{Fitted parameters}} \\
\cmidrule(lr){2-5}\cmidrule(lr){6-12}
\textbf{Taurus} 
& $d$ [pc] & $M$ [$M_\odot$] & $R_{\rm eff}$ [pc] & $R_{\rm A}$ [pc]
& $\log_{10} D_0$ & $\eta$ & $A$ & $a_2$ & $E_c$ [GeV] & $\chi_{\rm log}^2/\nu$ & $\sigma_{\Delta\log F}$ \\
\midrule
\textbf{Full cloud} & 147 & $1.7\times 10^4$ & 17.3 & 17.3 & 26.64 & 0.90 & 0.61 & \multicolumn{3}{c}{\textsc{Galprop spectrum}} & 0.054 \\
Taurus-c1   & 140 & $2.8\times 10^3$ & 1.10 & 0.24 & 26.25 & 0.50 & 1.45 & 3.00 & 2.69 & 1.78 & 0.14\\
Taurus-c2   & 150 & $4.2\times 10^2$ & 0.59 & 0.25 & 26.14 & 0.46 & 1.81 & 3.00 & 2.62 & 2.74 & 0.22 \\
Taurus-c3   & 140 & $1.9\times 10^3$ & 1.10 & 0.24 & 26.57 & 0.99 & 3.22 & 2.98 & 2.61 & 5.13 & 0.19 \\
Taurus-c4   & 140 & $2.4\times 10^3$ & 1.40 & 0.24 & 26.27 & 0.50 & 2.09 & 2.99 & 2.62 & 5.10 & 0.91 \\
\bottomrule
\vspace{0.1cm}
\end{tabular}
\caption{Physical properties and fitted CR-transport parameters for the Taurus molecular cloud and selected $\gamma$-ray-detected clumps embedded within the cloud complex. The parent Taurus cloud is  
fitted with $\gamma$-ray spectral measurements \citep{Neronov2017A&A}. The listed substructures are $\gamma$-ray-detected clumps from \citet{Yang2023NatAs}. For the parent Taurus cloud, distances and global properties are adopted from \citet{Cahlon2024ApJ}. For the clumps, we adopt the properties of \citet{Yang2023NatAs} to enable a self-consistent comparison with their $\gamma$-ray analysis to assess spatial variations in the effective propagation parameters within each cloud. For these clumps we define an effective radius $R_{\rm eff} = \sqrt{ab}$, the geometric mean of the reported half-major and half-minor axes $a$ and $b$ (approximating each region as an ellipse). $\gamma$-ray emissivity spectra were provided by Ruizhi Yang (private communication), obtained within an aperture region of radius $R_{\rm A}$ (here, we consider this to be the approximate smallest resolvable scale with an instrument like \textit{Fermi}-LAT for clumps, or the approximate size of the analysis region for Taurus in \citealt{Neronov2017A&A}). For clumps, we report an effective relative goodness-of-fit log-space statistic $\chi_{\rm log}^2 = \sum_i [(\log_{10} F_i - \log_{10} M_i)/\sigma_{{\rm log},i}]^2$ where $F_i$ and $M_i$ are the measured and model-predicted band fluxes, respectively, and $\sigma_{{\rm log},i}$ is obtained from the asymmetric flux uncertainties mapped into log-space and combined in quadrature with an energy-dependent floor. We do not show such a fit parameter for clouds, as the higher complexity of the boundary properties make it less meaningful. For clump fits, $N = 6$ and $\nu = 1$ (Stage 2 fits only $\log D_0$ and $\eta$ with other parameters frozen from Stage 1; see Appendix~\ref{sec:fitting_procedure}), which is small. Therefore we also report the RMS of $\Delta\log_{10}F$ in the final column (in dex) as an indicative dispersion metric. 
}
\label{tab:fit_results}
\end{table*}

We aim to fit our parameterized CR propagation model to the $\gamma$-ray {spectrum 
of the} Taurus molecular cloud, as reported by \citet{Neronov2017A&A}. To do this, 
we use a forward-modeling approach based on our CR transport framework. Model predictions are generated on a dense $\gamma$-ray energy grid where, 
for a given parameter set, the transport solver produces a proton distribution $n_\mathrm{p}(E_\mathrm{p},r)$ inside a spherical cloud model. The physical 
properties of the cloud are adopted from~\cite{Cahlon2024ApJ} (see Table~\ref{tab:fit_results}), and 
we impose a fixed incident CR proton spectrum from \textsc{Galprop}, representing the local interstellar CR population incident upon the cloud. In practice, the \textsc{Galprop} boundary condition specifies the CR proton number density at the cloud edge (outer radius $R_{\rm eff}$) as a function of energy, removing additional nuisance freedom in the spectral shape of the boundary. The exact normalization may deviate from the \textsc{Galprop} prediction, 
given energy-independent effects that we do not put focus on in this work (e.g. magnetic mirroring/focusing, external variations in the CR sea level, and geometric differences between exact modeled vs. observed aperture sizes, and geometric discrepancies arising from our 3-dimensional spherical model being fit to 2-dimensional plane-of-sky data). We absorb these into a free fitting parameter, $A$, which is expected to be of order unity. 
This ensures that the fit primarily constrains the transport parameters internal to the cloud. 

From our CR proton distribution, we compute the $\pi^0$-decay $\gamma$-ray emissivity per H atom, $q_\gamma(E)$, integrated over approximately the same physical aperture as the data, with characteristic size $R_{\rm A}$ reported in Table~\ref{tab:fit_results}. This is then ``observed'' in the same way as the data by top-hat averaging the model SED over each bin (sampling uniformly in $\log E$ within the bin) to ensure that comparisons are made consistently for wide bins and curved spectra. We detail our fitting procedure in Appendix~\ref{sec:fitting_procedure}, {with the} results reported in Table~\ref{tab:fit_results}.

\subsubsection{Application to clumps as embedded substructures}
\label{sec:clumps_application}

We also conduct a fit to the observed $\gamma$-ray spectra of clumps resolved within the Taurus molecular cloud. 
These are drawn from the sample of clumps in Taurus analyzed by 
\citet{Yang2023NatAs}. This is not intended as a precision measurement of 
the transport parameters for individual clumps. The $\gamma$-ray spectra at clump scales are noise-limited and the parameter space contains genuine degeneracies between the assumed boundary spectrum and the transport response. In this regime, formally optimizing a multi-parameter model can produce apparently well-constrained best-fit values that are not uniquely informative.\footnote{\citet{Yang2023NatAs} mitigated this issue by fitting the spectrum from all clouds in the sample combined. This stabilized the inferred boundary shape.} Instead, our aim in this illustrative demonstration is to set out an operational procedure to outline how future datasets could be used to conduct precision tests of CR transport with multi-wavelength anchors. We use the forward-modeling fit as a controlled way to generate self-consistent model realizations that reproduce the observed SEDs within uncertainties. This allows the associated multi-wavelength X-ray and ionization signatures in our two-zone transport scenario can be explored, and the prospects to detect such signatures to be considered in a context that is representative of 
that in which real data may be obtained. In this sense, the fitted parameters should be interpreted as indicative configurations of CR transport in molecular clumps in general, rather than definitive inferences of the specific target systems. They allow us to demonstrate the kinds of spectral and morphological signatures that could arise if transport is suppressed or enhanced in dense substructures, and identify which clumps/regions would be most promising targets for future, higher-S/N tests of CR transport (e.g. with deeper $\gamma$-ray data and complementary multi-wavelength tracers).

The general fitting procedure that we adopt for clumps follows that 
for the overall Taurus cloud, except that 
we now adopt a different boundary condition. This is because 
the 
CR spectrum at the cloud edge is expected to be substantially distorted compared to the LIS we model with \textsc{Galprop}. The propagation of CRs through the parent cloud in which the target 
clumps are embedded will modify the spectrum, especially if substantial scattering/diffusion and energy-dependent attenuation arises along the gas column. We therefore adopt a model-agnostic approach 
and invoke a smooth broken power-law as a boundary condition to be fitted in our analysis. This allows the boundary-shape parameters to be regularized using priors anchored to the CR spectrum in the enveloping molecular cloud, so that the boundary model does not absorb the signatures of transport suppression that we aim to illustrate. 
The boundary spectrum for species $i$ is written as: 
\begin{equation}
n_i(E_{i}) \;=\; n_{0,\rm{CR}} \;\! A\left(\frac{E_{i}}{E_0}\right)^{-a_1}
\left[1+\left(\frac{E_{i}}{E_c}\right)^{a_2-a_1}\right]^{-1},
\end{equation}
with $n_{0,\rm{CR}}$ set at a level comparable to the \textsc{Galprop} boundary spectrum (1.0 GeV$^{-1}$ cm$^{-3}$) and with a fixed low-energy spectral index $a_1$, and free parameters controlling the high-energy index $a_2$, break energy $E_c$, to be determined in the fit. The detailed fitting procedure is set out in Appendix~\ref{sec:fitting_procedure}.

Our results for each of the clumps are reported in Table~\ref{tab:fit_results}, where 
we show the same parameters as fitted of the overall Taurus cloud, and additionally the fitted spectral parameters for the boundary condition, which illustrates that the variation seen in the spectra between the different targets is 
achieved with only marginal variation of the boundary flux. 

\subsubsection{$\gamma$-ray signatures of CR transport}
\label{sec:taurus_diff_interpretation}

\begin{figure}
    \centering
    \includegraphics[width=\columnwidth]{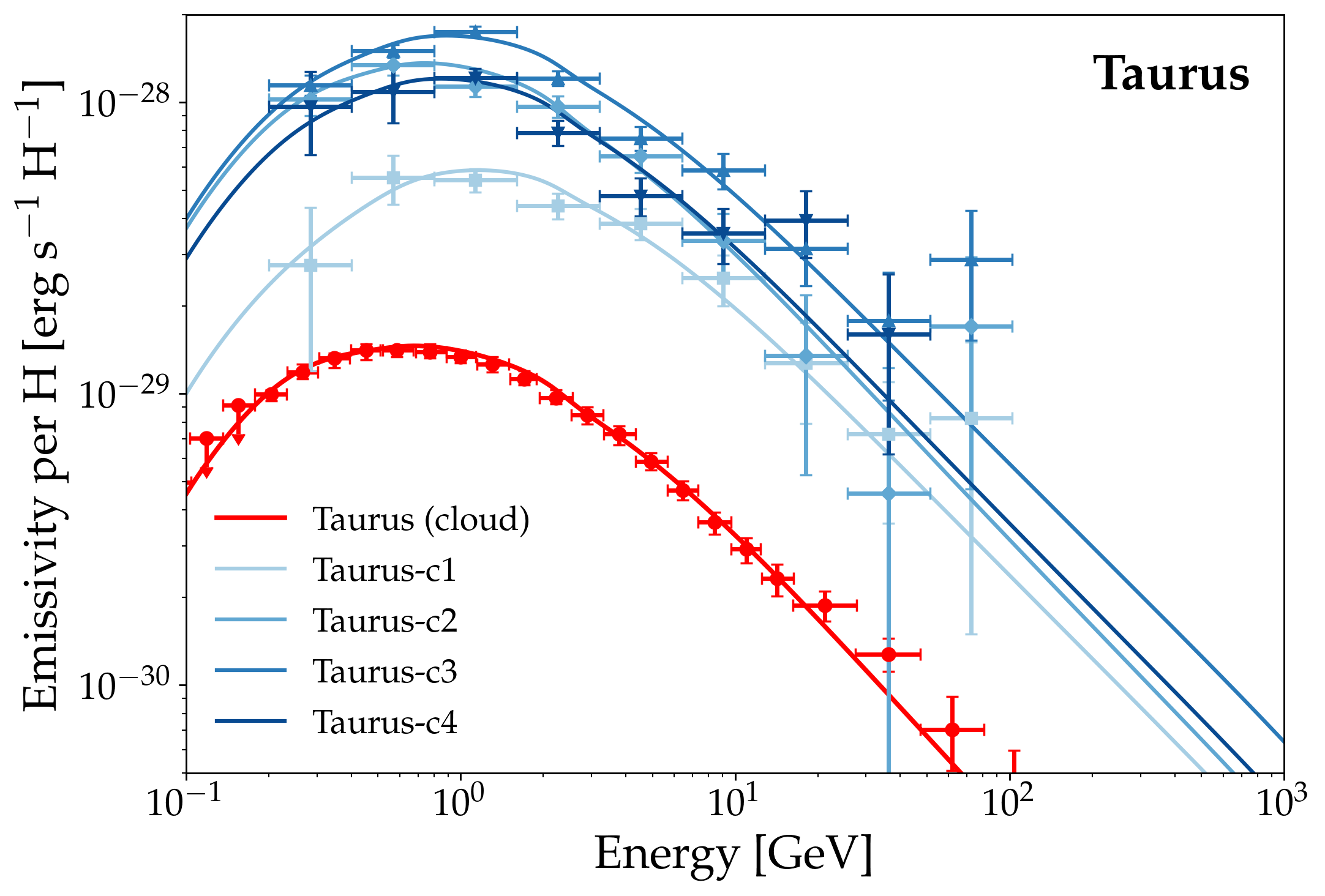}
    \caption{Pion-decay $\gamma$-ray emissivity per H atom for the Taurus molecular cloud (red; with over-plotted data points from~\citealt{Neronov2017A&A}) and representative embedded clumps (blue; data from~\citealt{Yang2023NatAs}). Points show the binned \textit{Fermi}-LAT emissivities and quoted uncertainties, while solid curves show the corresponding best-fit spectra from our CR-transport model adopted for each target (Table~\ref{tab:fit_results}). Relative to the cloud-scale spectrum, the clumps show both a higher specific emissivity and a stronger low-energy turnover/suppression below a few $10$~GeV; in a strong-scattering (slow-diffusion) interpretation of CR transport through the region, this requires diffusion coefficients at 1~GeV suppressed by up to $\sim 300$ compared to the canonical ISM value.}
    \label{fig:clump_gamma_fit} 
\end{figure}

We plot the fitted pion-decay $\gamma$-ray emissivities per H atom for the Taurus cloud and representative embedded clumps in Figure~\ref{fig:clump_gamma_fit}. Two robust empirical trends emerge. First, the clumps are systematically brighter per unit gas mass than the cloud-scale average, with specific emissivities higher by roughly an order of magnitude. Second, the clump spectra exhibit a markedly stronger low-energy turnover than the cloud-scale spectrum. Below a few-$10$~GeV the emissivity is increasingly suppressed, and the degree of suppression varies from clump to clump.

These spectral differences are already apparent by eye in the \textit{Fermi}-LAT points, independent of the details of the forward modeling. In our transport interpretation, reproducing the enhanced low-energy attenuation within the clump apertures requires substantially stronger scattering than on cloud scales. Consistent with the fit results in Table~\ref{tab:fit_results}, all clumps favor diffusion coefficients suppressed by at least $\gtrsim 2$~dex relative to a canonical ISM normalization at GeV energies, with the most strongly attenuated clumps requiring suppressions approaching $\sim 300$ at 1~GeV. The fits further indicate that the slow-transport region occupies a substantial fraction of each clump aperture, extending out to radii where the gas density remains a significant fraction of the peak clump density, rather than being confined only to the very densest core.

While these interpretations  
may be considered as a scenario where substantial CR scattering and coupling is arising throughout these dense structures, 
the clump-to-clump scatter in the turnover energy and suppression amplitude (a factor of a few across our sample) cautions against over-interpreting any single pair of best-fit parameters $(D_0,\eta)$ as a unique physical diffusion law. Given the present statistics and the limited set of apertures, the most robust conclusion is qualitative, i.e. that the data require appreciable low-energy attenuation of the incident CR spectrum before it reaches the central, densest regions of the clumps sampled by the \textit{Fermi}-LAT analysis.

If this behaviour is indeed caused by slow diffusion (strong scattering) in dense substructure, a key multi-wavelength implication is that the associated secondary electron population, and hence its non-thermal emission, should be significantly enhanced relative to LIS-irradiated expectations. In the following sections we therefore adopt the slow-diffusion interpretation as a working hypothesis and outline the signatures it predicts across complementary tracers, noting that tighter constraints on transport parameters will ultimately require joint fits across cloud and clump scales, multiple apertures, and improved $\gamma$-ray statistics.

\subsubsection{Prospects for multi-wavelength tests}
\label{sec:fitting_results}

\begin{figure}
    \centering
    \includegraphics[width=\columnwidth]{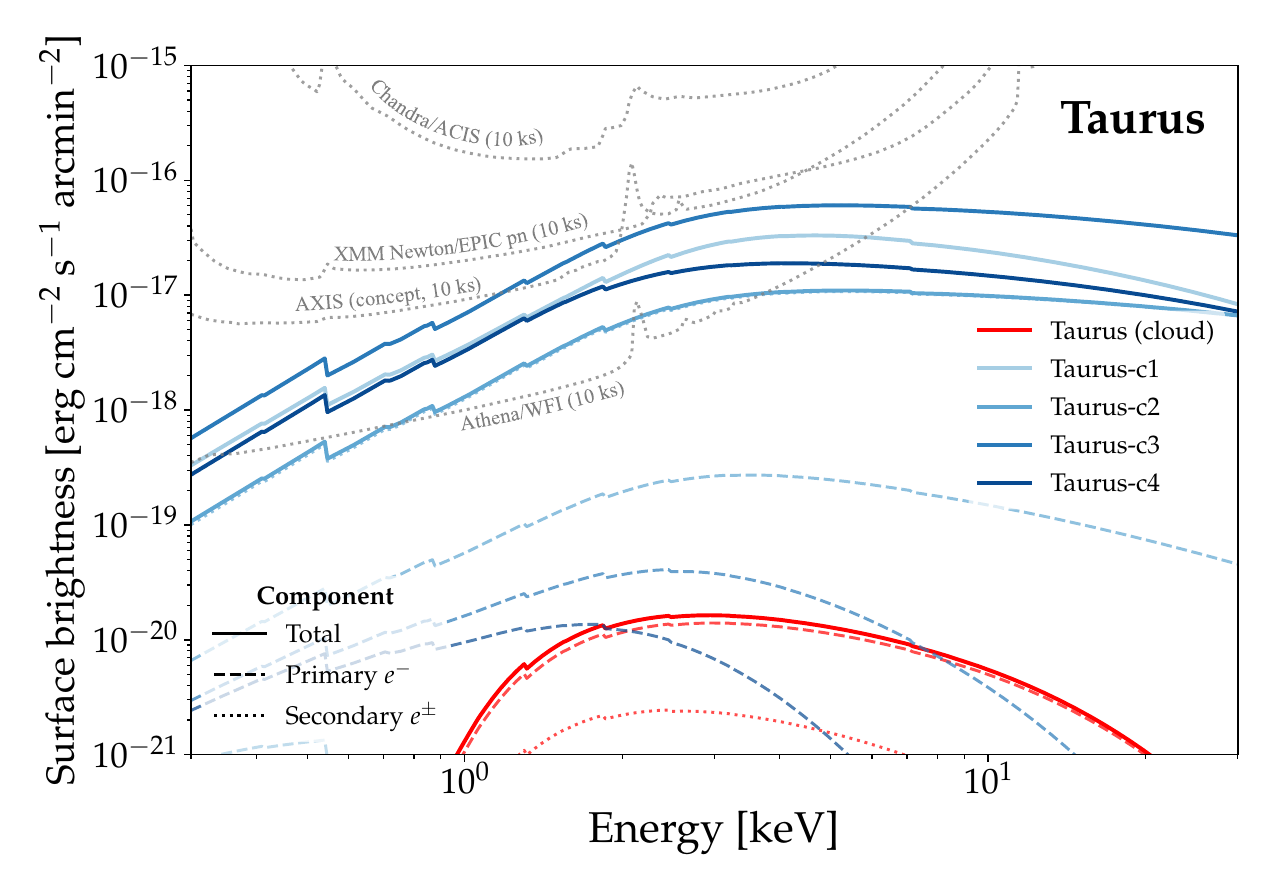} 
    \caption{{Predicted total absorbed X-ray surface-brightness spectra of synchrotron emission from CR electrons for the Taurus molecular cloud (red) and representative embedded clumps (blue) under the model parameter fits shown in Figure~\ref{fig:clump_gamma_fit}. Contributions from primary and secondary electrons are indicated using dashed and dotted lines, respectively, where the secondary component is dominated by 0.1-1 TeV electrons.} Gray dotted curves show illustrative, 
    counts-limited surface-brightness scalings (not full detectability forecasts) estimated by requiring a minimum of $N_{\min}=50$ source counts in one logarithmic energy bin ($\Delta\log_{10}E=0.2$) for an exposure time of 10 ks, with several current and proposed instruments (see Appendix~\ref{app:xray_sens} for details). As these clouds would be extended due to the close proximity of Taurus, the extraction area is capped at the instrument field of view, and we note that our background-limited performance may be less favorable for extended emission than estimated here.}
    \label{fig:xray_predictions_taurus}
\end{figure}

\begin{figure*}
    \centering
\includegraphics[width=0.48\linewidth]{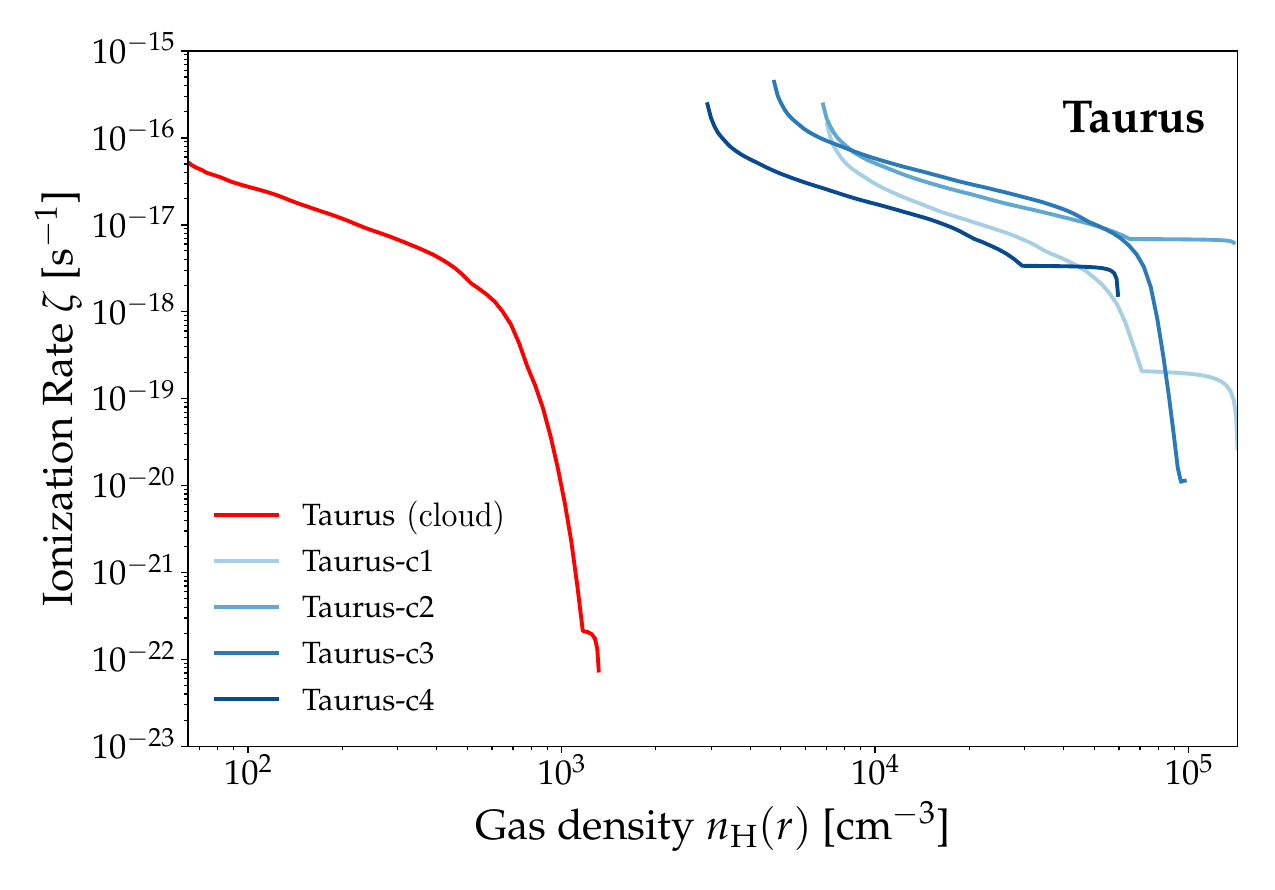} 
    \hspace{0.2cm} 
\includegraphics[width=0.48\linewidth]{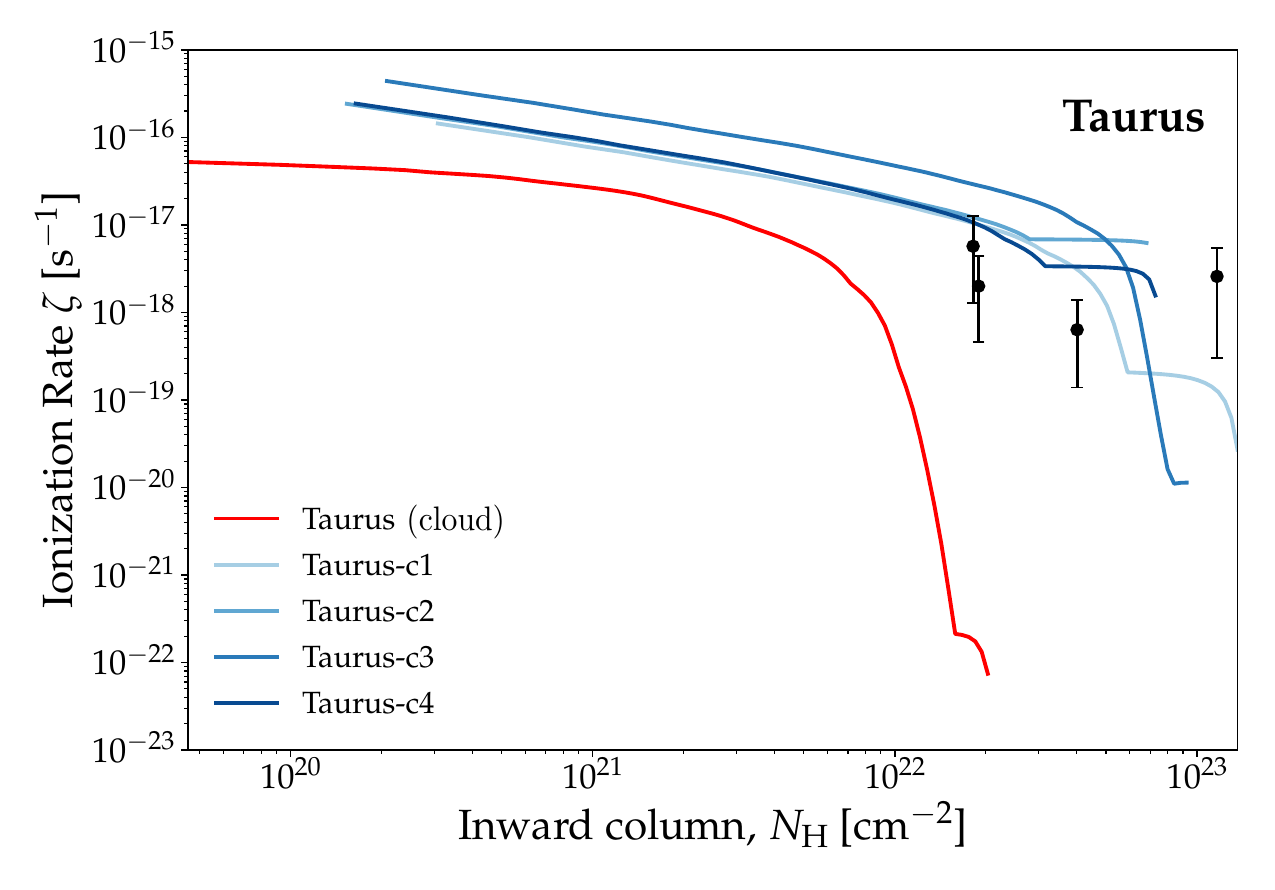}
    \caption{Predicted CR ionization rate per H atom, $\zeta$, as a function of (left) local gas number density, and (right) inward hydrogen column density the CRs propagate through, $N_{\rm H}$, for the Taurus molecular cloud (red) and the four embedded $\gamma$-ray--bright clumps (blue). The left abscissa is the 
    local density of the gas, while the right abscissa is the 
    radial column integrated from the cloud surface to depth $r$. This ensures that each curve traces the decline of $\zeta$ into the shielded interior as low-energy primary CRs are progressively attenuated. Black points on the right plot show representative ionization-rate constraints for Taurus dense cores from \citep{Redaelli2025A&A}, converted from the commonly quoted per-$\mathrm{H}_2$ convention to a per-H rate for direct comparison.}
    \label{fig:ionization_results}
\end{figure*}

If the low-energy suppression seen in clump-scale $\gamma$-ray spectra is driven by slow diffusion (strong scattering) in dense substructures,  
the implications would 
extend beyond the GeV band. 
The impacts of slow diffusion would manifest throughout the electromagnetic 
emission from the penetrating CRs. In particular,  enhanced confinement increases the hadronic interaction yield and boosts the resulting secondary electron 
population. This implies potentially detectable non-thermal emission and ionization signatures that can be used as independent tests of the transport interpretation. 
Here, we consider $\gamma$-ray-forward-model illustrative realizations for the Taurus cloud and representative clumps to generate self-consistent predictions for synchrotron X-rays (Figure~\ref{fig:xray_predictions_taurus}) and CR ionization rate as a function of column depth (Figure~\ref{fig:ionization_results}) as two complementary tracers. 

Figure~\ref{fig:xray_predictions_taurus} shows the 
predicted absorbed X-ray surface-brightness spectra for the parent Taurus cloud and the fitted clumps, demonstrating how the emission from the clumps is strongly dominated by the contribution from secondary CR electrons.  
{This 1-10 keV synchrotron emission from the clumps is dominated by secondary electrons of energies of $\sim 0.1-1$ TeV, produced 
in situ by CR proton interactions with the dense gas.} 
To relate intrinsic emissivities to an observable quantity, we adopt a conservative line-of-sight absorbing column density of $N_{\rm H}=2.3\times10^{22}\,{\rm cm^{-2}}$ for all clumps and 
the Taurus cloud, representative of the upper end of relevant lines of sight columns to the Taurus--Perseus complexes as inferred from Planck maps of the region~\citep{Planck2014A&A}. From this, 
we impose absorption as a uniform foreground screen applied to all emitted photons to estimate the 
effect of interstellar attenuation of X-rays. We additionally adopt an extraction scale comparable to the clump $\gamma$-ray analysis region for each structure (their 
$R_{\rm A}$ in Table~\ref{tab:fit_results}), noting that Taurus is nearby and any diffuse signal is extended on the sky. 

To provide an order-of-magnitude feasibility reference for detecting the predicted extended X-ray emission, overlaid dotted curves show order-of-magnitude, counts-limited surface-brightness sensitivity scalings for {several current, future and representative prospective instrument capabilities} (\textit{Chandra}/ACIS-I, \textit{XMM-Newton}/EPIC-pn, {AXIS, as considered in the mission concept study and included here as the indicative performance of a possible future mission}, and \textit{Athena}/WFI) for a single 10 ks pointing, intended as a feasibility guide (see Appendix~\ref{app:xray_sens}). 
 In this framework, the clumps are generically predicted to be more promising than the cloud as targets for diffuse non-thermal X-ray emission due to their 
 higher density (boosting hadronic interaction rates and injection of secondary 
 CR electrons, which dominate the clump X-ray emission) and stronger magnetic fields than the more diffuse parent cloud. This motivates future deep, carefully background-controlled observations 
 as a direct test for the enhanced-secondary 
 consequence of a substantial slow-diffusion zone in the structures. 

The ionization-rate predictions in Figure~\ref{fig:ionization_results} provide 
a complementary probe of the low energy CR population. 
These CRs are most directly affected by attenuation and transport in dense gas, 
and are most sensitive to the size of any diffusive layer. We plot the total CR ionization rate $\zeta$ against local gas density on the left panel. This representation is useful because most observational constraints on $\zeta$ are not direct local measurements of $\zeta(r)$, but rather density- and chemistry-weighted inferences from tracers that preferentially sample particular regimes of $n_{\rm H}$. 
As a result, the same underlying $\zeta(r)$ profile can yield systematically different inferred ``characteristic'' ionization rates depending on whether the diagnostic is envelope-weighted (diffuse/translucent gas) or core-weighted (dense gas). On the right panel, we show the ionization rate in terms of the 
inward gas column experienced by the propagating CRs, $N_{\rm H}$, enabling a like-for-like comparison between cloud and clump configurations and a direct connection to observational inferences of $\zeta$ that are naturally tied to a characteristic column/density regime. 

In the $\gamma$-ray-fitted Taurus realizations, the parent cloud shows 
only modest variation with depth, 
while the clumps show substantially stronger suppression toward large columns. 
This reflects the increasing difficulty for low-energy CR primaries to penetrate to the densest interiors, 
and is consistent with the general trend that CR ionization rate decreases 
with gas column density~\citep[e.g.][]{Indriolo2012ApJ}. The overlayed $\zeta$ measurements (adapted from \citealt{Redaelli2025A&A}) illustrate that constraints on CR transport can already be placed using these trends, but they should be interpreted with care. In particular, any inferred $\zeta$ is effectively an emission- or absorption-weighted average over the density/column regime probed by the chosen tracer, rather than a direct measurement of the $\zeta(r)$ profile itself. 

\section{Discussion}
\label{sec:discussion}

\subsection{Observational prospects and interpretations}
\label{sec:model_parameters}

Taken together, our $\gamma$-ray spectra (anchoring the $\gtrsim$GeV proton population), secondary-driven synchrotron X-rays (anchoring the secondary electron yield), and ionization constraints (anchoring the sub-GeV CR component) provide a route to breaking degeneracies in CR-transport modeling that are difficult to resolve with $\gamma$-rays alone. In particular, they separate boundary-shape effects from transport responses, test whether a distinct diffusion zone is required, and assess whether CR scattering/coupling (``slow diffusion'') is enhanced in dense ISM substructures relative to the wider ISM. 

From an observational standpoint, the natural progression is hierarchical. $\gamma$-rays provide the most direct and currently accessible handle on CR attenuation in dense gas. They identify the clumps where a low-energy deficit is present, and provide a constraint on the extent of the affected region (parameterized in this framework by $\eta$). X-rays and ionization then act as complementary follow-up probes. X-rays test whether the inferred suppression is accompanied by the increased secondary yield expected in diffusion-dominated propagation (and therefore constrain the degree of slowing/scattering). 
Ionization probes the sub-GeV CR population and its penetration into envelope versus core material. 

Observations that target the same set of well-characterized clumps with (i) deeper $\gamma$-ray measurements and multiple apertures, (ii) sensitive diffuse X-ray imaging optimized for extended low-surface-brightness emission, and (iii) ionization-sensitive tracers spanning envelope- and core-weighted regimes, offer a practical route to conduct robust tests of CR transport in dense gas with current or near-future facilities. In particular, 
clumps presenting stronger low-energy $\gamma$-ray suppression should show higher secondary signatures (higher non-thermal surface brightness) alongside systematically reduced core-weighted ionization rates relative to envelope-weighted inferences (potentially discernible from the choice of molecular ion tracer). This provides a falsifiable test of slow CR diffusion in dense ISM substructure. In the following subsections we discuss these diagnostics in more depth, outline target-selection strategies, and highlight key interpretation caveats in each waveband.

\subsubsection{$\gamma$-ray suppression maps}
\label{sec:gamma_suppression}

The multi-wavelength signatures discussed in this work are strongest when CR propagation through dense gas is diffusion-dominated and the effective traversed material is large. This is favored in compact, high-column structures (typically also higher-density and higher magnetic field strength), where increased magnetic scattering prolongs residence times, strengthens the attenuation of low-energy primaries, and boosts secondary production. 
This qualitative picture holds across the three tracers considered here, although the observational signatures (and therefore the target-selection criteria) differs by band. In $\gamma$-rays, increased CR coupling in dense gas 
appears most directly as a low-energy deficit (spectral curvature) below $\sim$100~GeV relative to the ballistic, free-penetration expectation. This deficit increases with either a more extended diffusion zone (larger $\eta$) and/or a smaller diffusion normalization $D_0$, but these effects are largely degenerate in $\gamma$-ray spectra alone. Despite this, a pronounced low-energy turnover remains one of the clearest signatures of slow diffusion in dense substructure.

\begin{figure}
    \centering
    \includegraphics[width=\columnwidth]{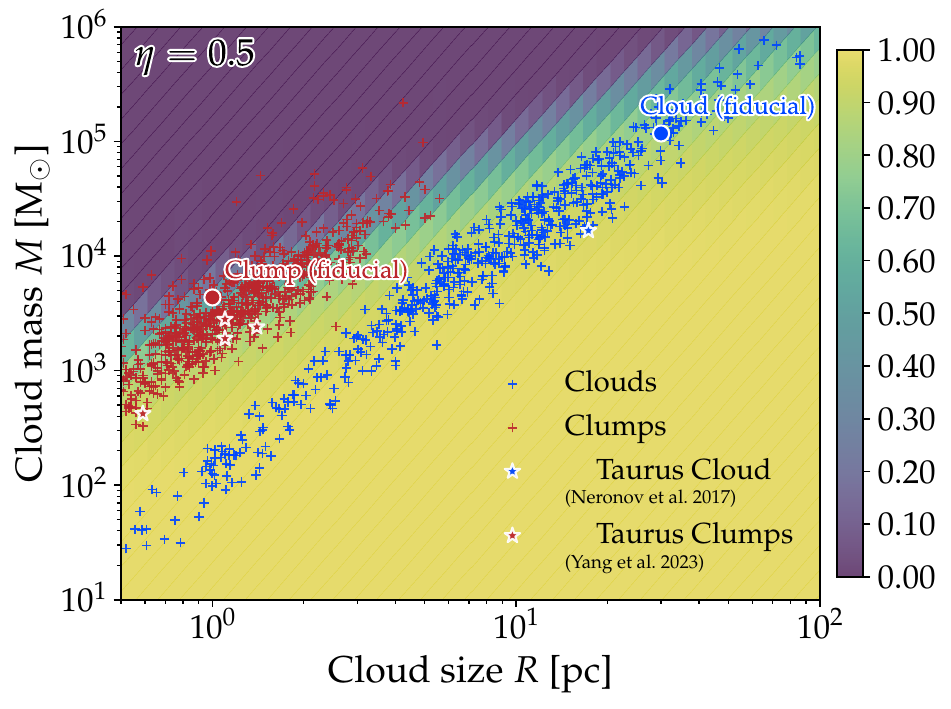}
    \caption{
    Mass-size map of the predicted low-energy $\gamma$-ray deficit in molecular structures for the fiducial two-zone CR-transport model with transition parameter $\eta=0.5$ (labeled), and diffusion coefficient normalization of $3\times 10^{26}$ cm$^2$ s$^{-1}$ set as that adopted for our fiducial clump model. Colors show the degree of deficit at $E_\gamma=1$~GeV,
where a value of 1 (yellow) indicates no suppression compared to a ballistic propagation scenario, while smaller values indicate a stronger low-energy deficit. Crosses indicate representative Galactic molecular clouds (blue) and dense clumps (red), with data obtained from~\cite{Urquhart2014MNRAS, Chen2020MNRAS}. Open circles mark the fiducial cloud and clump models used for the parameter studies. Star symbols highlight the Taurus cloud and the $\gamma$-ray--detected Taurus clumps used in the illustrative application in this work (section~\ref{sec:illustrative_applications}).}
    \label{fig:suppression_map}
\end{figure}

In practice, our ability to detect and characterize the GeV-band deficit 
is expected to be the restrictive bottleneck in selecting promising laboratories and building a sizable target sample. 
The overall detectability of molecular clouds in $\gamma$-rays is set by mass and distance, with nearby massive complexes offering the best prospects for detection \citep[e.g.][]{Aharonian2020PhRvD, Peron2022A&A}. However, establishing a GeV emission deficit also requires 
sufficient angular resolution and photon statistics to separate clump-scale emission from the parent-cloud envelope. 
$\gamma$-ray observations have lower angular resolution compared to other wavelengths, so this additional requirement typically limits viable targets to the nearest cloud complexes. 
Within these complexes, the cloud structure must be favorable. For fixed $(D_0,\eta)$, more compact, higher-column configurations produce stronger deficits because CRs encounter larger columns over shorter physical scales. Figure~\ref{fig:suppression_map} illustrates this dependence by showing the predicted 1~GeV deficit across the mass--size plane for our fiducial transport model. The fiducial cloud/clump configurations and the Taurus application are annotated for context. The map indicates that strong suppression is expected primarily on clump scales, and that the parameters adopted here are conservative given that suppression is already inferred for the Taurus clumps (red stars). The over-plotted catalog clouds and clumps \citep{Urquhart2014MNRAS, Chen2020MNRAS} therefore provide a natural starting point for identifying additional laboratories where GeV suppression (and hence strong coupling) is most likely, subject to the detectability requirements above.

\subsubsection{X-ray luminosity and attenuation}
\label{sec:xray_and_attenuation}

In our framework, secondary-driven synchrotron X-ray emission is expected to accompany all clumps that show GeV-band suppression in their $\gamma$-ray spectra. 
This is a general consequence of strong CR coupling in dense ISM substructures. 
This means that, if a cloud complex is close enough for a GeV deficit to be established, X-rays do not impose an additional astrophysical selection on the target sample.
 Every clump with a clear GeV suppression is a relevant X-ray test case. 
 The practical limitations are observational, i.e. instrument sensitivity and extended-source background control. This is because the signal may be spatially extended, and strongly attenuated at soft energies by photoelectric absorption. 
 Our surface-brightness and sensitivity estimates in Figure~\ref{fig:xray_predictions_taurus} indicate that current facilities 
 are unlikely to access even the brightest subset of GeV-selected clumps. 
This is consistent with the broader observational landscape. 
Similar dense structures in the vicinity of supernova remnants, where the incident CR flux should be substantially higher than for our targets, have also yielded non-detections with longer integration times on current facilities \citep{Tsuji2024ApJ}. 
Next-generation X-ray missions are expected to reach the sensitivities required to detect hard-band emission in the most favorable cases and may expand the accessible sample beyond Taurus to other nearby cloud complexes. If future observations of GeV-selected targets do not reveal the expected emission, this provides a crucial test. Stringent X-ray upper limits for GeV-suppressed clumps would directly falsify that GeV suppression is due to diffusion-dominated transport, especially if the clump magnetic field is well characterized. 

 Beyond providing an independent verification channel, X-rays also make the $\gamma$-ray measurements more informative. They provide scope to break the $(D_0,\eta)$ degeneracy in our framework. The GeV-band deficit is comparably sensitive to the severity of CR slowing, and to the spatial extent of any diffusion-dominated region. A smaller $D_0$ and a larger $\eta$ can both increase the effective gas column CRs traverse, and produce the same the GeV-band suppression. By contrast, the secondary synchrotron emissivity is set primarily by the interaction rate and residence time of CRs in dense gas. It is therefore expected to respond more strongly to changes in $D_0$ (which directly controls the CR confinement time, hence secondary injection) than to modest changes in $\eta$ (which mainly redistributes \textit{where} within a structure diffusion dominates and secondaries are injected). For a measured GeV suppression, X-ray detections or upper limits bound 
 the allowed secondary yield and restrict $D_0$ (or, in extreme cases, rule out diffusion-dominated interpretations). This is not possible with $\gamma$-rays alone. Joint 
 constraints from the $\gamma$-ray spectrum and hard X-ray surface brightness can therefore 
offer a practical pathway to decouple ``how slow'' diffusion is from ``how far'' the diffusion zone extends. $\gamma$-rays localize the suppression, while X-rays test whether the expected level of scattering and secondary production in a given transport theory is physically realized. 

\subsubsection{Ionization rates and chemical tracers}
\label{sec:ionization_strategies}

Measurements of the CR ionization rate, $\zeta$, provide an important additional constraint in our framework. 
They probe the sub-GeV CR population that is largely invisible to $\gamma$-rays and add new 
leverage on CR penetration into dense gas. 
In the same way that hard X-rays test whether GeV-selected clumps show the enhanced secondary yield expected for diffusion-dominated transport, $\zeta$ tests whether low-energy CRs are excluded from the clump interior relative to the envelope. 
If CRs propagate diffusively, the qualitative expectation is that clumps with stronger GeV suppression should show a stronger contrast between envelope-weighted and core-weighted ionization tracers. The envelope remains comparatively highly ionized because CRs experience limited attenuation at low column densities, while core-weighted tracers infer reduced $\zeta$ where the lowest-energy CRs primaries are most strongly attenuated. Previously reported anti-correlations between inferred $\zeta$ and gas column density 
into clouds~\citep{Albertsson2018ApJ} would be stronger under diffusion-dominated propagation, and markedly weaker if propagation were closer to ballistic.  Depending on the relative importance of secondary electrons at large columns (as independently tested by X-rays), the innermost trend can flatten or partially recover, so the most informative constraint is the comparison between envelope- and core-weighted diagnostics, rather than any single characteristic value of $\zeta$. 

Separating envelope and clump CR ionization observationally is possible because inferred $\zeta$ values are tracer-weighted averages over the regions where the relevant chemistry operates and where the observed transition is effectively excited. 
Different tracers (and even different transitions of the same species) sample different density/column regimes. Diffuse/translucent-gas diagnostics, for example absorption-line ions such as ArH$^+$, OH$^+$, H$_2$O$^+$ and H$_3$O$^{+}$ in suitable environments \citep[e.g.][]{Neufeld2017ApJ}, preferentially probe outer layers where attenuation is minimal and $\zeta$ is expected to be highest. In contrast, commonly used dense-gas tracers in sub-mm emission, such as low-$J$ lines of HCO$^+$, N$_2$H$^+$, and related deuterated ions \citep[e.g.][]{Caselli1998ApJ, Luo2024A&A}, weight shielded interiors where $\zeta$ is most sensitive to the attenuation of sub-GeV primaries and (in diffusion-dominated scenarios) to any secondary-driven ``floor'' at high columns (cf.\ Figure~\ref{fig:change_eta}).
The presence or absence of X-ray synchrotron emission and GeV suppression in the same regions can help discriminate whether suppressed primaries or secondary injection dominates the inferred ionization. 

{Recent work has further emphasized that CR attenuation can reshape abundance profiles and bias simple constant-$\zeta$ interpretations. \citealt{Roy2026A&A} proposed chemical ratios that provide more robust calibrators of $\zeta$  
in dense, UV-shielded gas, 
while 
\citet{Latrille2025A&A}  
demonstrated through post-processed cloud simulations that CR propagation through dense material can significantly affect both deuterated and non-deuterated tracers. These studies support the need to compare ionization models to the density and column regimes actually sampled by each tracer.} 
In this context, the most robust strategy is to combine at least one envelope-weighted and one core-weighted ionization diagnostic for the same GeV-selected targets, and to compare them to model predictions matched to the density/column regime actually sampled, rather than comparing a single inferred $\zeta$ value to an entire radial profile.

\subsubsection{Complementary multiwavelength diagnostics}
\label{sec:other_wavelenghts}

This work has focused on $\gamma$-rays, X-ray synchrotron emission from secondary
electrons, and CR-driven ionization signatures as the most promising
observables for isolating CR transport physics in dense molecular environments.
These probes are not exhaustive, and observations at other wavelengths can provide
complementary constraints and additional consistency checks.
In particular, radio synchrotron emission is a natural counterpart to the X-ray
synchrotron component considered here
\citep[e.g.][]{Protheroe2008MNRAS,Rodriguez2013ApJ}
and may become accessible with next-generation facilities such as the Square
Kilometer Array Observatory (SKAO; \citealt{Braun2015aska}). In principle, SKAO
could detect diffuse synchrotron emission from nearby Galactic molecular cloud
cores, particularly at low radio frequencies \citep{Dickinson2015aska,Padovani2018A&A}.
Moreover, radio polarization measurements may offer additional information on
magnetic-field structure and turbulence within clumps and cores \citep{Dickinson2015aska}.

While such emission would provide useful insights about the physical conditions within 
molecular clouds, radio emission is not a clean diagnostic
of CR propagation. {At radio frequencies,} the emission is typically dominated by
relatively low-energy electrons, and the synchrotron brightness depends strongly on both the
magnetic-field strength and its local structure. As a result, spatial variations in magnetic
structure can masquerade as changes in particle penetration or transport, complicating attempts
to isolate CR transport physics using intensity information alone. 
By contrast, ionization-rate
constraints probe CRs through their interaction rate with the gas and are comparatively less
sensitive to small-scale magnetic-field structure, providing a cleaner signature of CR
penetration into dense clumps at similar CR energies. Used together, radio synchrotron and
ionization constraints could nonetheless provide useful additional leverage on the underlying
CR composition (leptonic vs. hadronic) and help build a more complete physical picture of dense clump environments,
including possible indications of the microphysical origins of slow CR transport. 

\subsection{Further remarks}
\label{sec:further_remarks}

Our results explore the possibility that CRs undergo substantial scattering in dense substructure. Such a scenario would imply non-negligible coupling and energy/momentum exchange between CRs and cold gas phase, at least on the scales of clumps. The inferred GeV-band suppression in $\gamma$-rays is 
naturally interpreted as hindered penetration of low-energy CRs into high density gas. This points to strong engagement of CRs with dense material over the scales probed here. If confirmed across a broader sample, it would suggest that CR coupling in dense ISM substructures can be stronger than is often considered for partially neutral or multi-phase media in galaxies. This would be tension with the common theoretical expectation that ion-neutral damping suppresses self-confinement and leads to effective decoupling \citep{Everett2011ApJ, Plotnikov2021ApJ}.

Understanding how suppressed diffusion and large effective filling factors we obtained could arise micro-physically is challenging. In particular, a key open question is then geometric - and whether the coupling is genuinely volume-filling throughout clumps, or concentrated in diffusion-dominated outer layers (a ``skin''), with only limited CR interaction deeper in. The multi-wavelength diagnostics discussed in Section~\ref{sec:model_parameters} offer a practical route to distinguish these possibilities, since the joint pattern of GeV suppression, hard X-ray secondaries, and envelope-versus-core ionization constraints directly probe where CRs are excluded, where their attenuation accumulates, and where secondaries dominate. Another plausible interpretation is that transport is controlled by intermittency. Very slow diffusion could operate in localized magnetic traps or strongly scattering sub-regions embedded within a more weakly scattering volume (i.e. a ``raisin-bread'' picture). These could be caused by the magnetic field topology, or even ionized embedded cavities due to enshrouded pre-main-sequence stars, brown dwarfs or compact objects. Indeed, in the Taurus complex we considered in this work, 
pre-main-sequence stars and brown dwarfs have been reported to be identified throughout the
dense clump structures~\citep{Gudel2007A&A, Grosso2007A&A}. These can produce persistent and flaring keV emission, carving-out localized regions of higher ionization, where ion-neutral damping would be reduced.   
In such an inhomogenous medium, the effective large-scale transport can be regulated by the slowest regions, with the overall diffusion more closely related to a harmonic mean of the spatially varying diffusion coefficient \citep{Ewart2025arXiv}. This naturally motivates transport prescriptions that go beyond a single uniform $D_0$ and instead treat the dense ISM as an intermittent medium where rare slow-diffusion structures dominate the residence time \citep[see e.g.][]{Liang2025MNRAS}. In this view, the ``large'' effective diffusion zone inferred from the $\gamma$-ray data could reflect a region under the influence of intermittent transport bottlenecks that prolong confinement without requiring uniformly slow diffusion everywhere.

Interpreting GeV suppression requires careful control of gas templates and unresolved substructure. Biases in the emissivity-per-H conversion can mimic spectral curvature or generate spurious clump--envelope differences. However, if extensive slow CR diffusion on clump scales is confirmed, it would have several broader implications. First, CR exclusion from dense interiors would imply a local ``CR hole'' and hence a CR pressure gradient at the cloud/clump interface. This can excite MHD fluctuations and potentially modify local transport to drive dynamical feedback on the gas \citep[e.g.][]{Bustard2021ApJ}. Second, stronger coupling increases the potential importance of CR heating and ionization for the thermal balance and chemistry of dense gas, with consequences for feedback and star-formation regulation \citep{Papadopoulos2010ApJ, Owen2021ApJ}. Even if slow diffusion primarily reduces the low-energy CR density in the deepest clump interiors, increased residence times and boosted interactions in the coupled regions can still boost the integrated secondary yield and local energy deposition where CRs do interact. In this sense, the same transport physics that produces GeV suppression can also amplify CR-driven ionization/heating in parts of the structure that remain coupled. Finally, stronger CR coupling behavior would bear directly on phase-dependent CR transport prescriptions used in galaxy/ISM simulations, where rapid CR transport through cold gas (or explicit decoupling) is often adopted \citep[e.g.][]{Farber2018ApJ, Armillotta2021ApJ, Habegger2024ApJ, Sike2025ApJ}. If CR coupling to dense gas is stronger than assumed, new sub-grid treatments may be required. Ultimately, establishing whether strong coupling is common, and identifying the microphysical origin of the required scattering, will require joint constraints on CR attenuation, secondaries, and gas/magnetic structure across a broader set of nearby cloud complexes.

\section{Summary and conclusions}
\label{sec:summary}

In this work, we investigated how the penetration and coupling of the ambient Galactic CRs into dense molecular structures can be constrained using a combined set of diagnostics: pion-decay $\gamma$-rays (anchoring the $\gtrsim$GeV proton population), secondary-driven synchrotron X-rays (anchoring the secondary electron yield), and CR ionization rates (anchoring the sub-GeV component). We developed a simple, physically-motivated transport framework spanning three limiting behaviors: (i) ballistic propagation, (ii) fully diffusive propagation, and (iii) a two-zone configuration where an outer diffusion-dominated layer surrounds an interior core where CRs propagate ballistically. 

We confirmed earlier interpretations~\citep[e.g.][]{Yang2023NatAs, Jiang2025MNRAS} that $\gamma$-ray suppression arises naturally once CR propagation through compact, high-column clumps becomes diffusion-dominated. Enhanced magnetic scattering in such structures prolongs residence times and increases the effective attenuation of CRs to pp interactions accumulated before CRs reach the core. This produces a marked low-energy deficit (spectral curvature) of the pion-decay emissivity below $\sim 10^2$~GeV relative to a ballistic, free-penetration expectation. The same mechanism is expected to be weak for more extended, lower-column cloud-scale structures. Interpreting strong suppression in terms of diffusion-dominated transport implies that the degree of scattering in dense substructure is substantial. If confirmed across a broader sample, this would be in tension with the common expectation that CRs couple poorly to dense interstellar gas. 

Any diffusion-dominated region that produces a GeV deficit also generically boosts the secondary yield and thus the intrinsic synchrotron emissivity from secondary CR electrons within clumps. After applying photoelectric absorption, soft X-rays can be strongly attenuated, but the hard-band separation between ballistic and diffusion-dominated configurations remains, making hard X-rays an independent test of whether GeV-selected clumps present enhanced secondary production required by slow-diffusion interpretations. 

Ionization rates provide complementary leverage on the sub-GeV CR population that is largely invisible to $\gamma$-rays and particularly sensitive to whether low-energy CRs penetrate into dense interiors. Our models predict that $\zeta$ can vary substantially with depth into dense structures, so the most informative strategy is to compare envelope-weighted and core-weighted ionization diagnostics for the same GeV-selected targets, rather than relying on a single characteristic $\zeta$ value. In diffusion-dominated interpretations, stronger GeV suppression should correspond to a stronger envelope--core contrast in $\zeta$. 

Applying our framework to the Taurus cloud and its $\gamma$-ray detected clumps, we find that the observed clump spectra are naturally reproduced when slow diffusion operates over a substantial fraction of the clump volume, with best-fit parameters implying diffusion coefficients well below canonical ISM values. In this interpretation, the same transport configuration that produces GeV suppression predicts enhanced secondary signatures in hard X-rays and systematically reduced core-weighted ionization relative to envelope-weighted inferences, providing a clear set of falsifiable multi-wavelength tests.

From an observational perspective, 
a hierarchical strategy arises naturally to test 
for the presence of slow CR diffusion in dense clumps. 
$\gamma$-rays provide the most direct handle on GeV-scale exclusion and identify the subset of nearby clumps. Indeed, a suppression is measurable with current facilities and sufficient angular resolution. Hard X-rays and ionization constraints then act as complementary follow-up probes that test (i) whether the required secondary yield is present and (ii) how deeply low-energy CRs penetrate into envelope versus core material. Next-generation X-ray missions are expected to be particularly important for pushing beyond current sensitivity limits for diffuse, absorbed, low-surface-brightness emission and for expanding the accessible sample beyond the nearest cloud complexes. 
More broadly, if strong clump-scale coupling is common, it would motivate revisiting phase-dependent CR transport prescriptions in which CRs are assumed to traverse cold dense gas efficiently or to decouple from it. Establishing whether the required scattering arises from pervasive turbulence, magnetic trapping/intermittency, or self-generated fluctuations will require extending this analysis to a broader set of nearby clouds with well-controlled gas templates and deep X-ray/ionization observations of $\gamma$-ray-selected clumps. 

\section*{Funding}
HPHN and STC acknowledge funding by the Helga-Kersten program supporting faculty-specific gender equality targets at Friedrich-Alexander University Erlangen-N\"{u}rnberg (FAU), which supported an internship at the Erlangen Centre for Astroparticle Physics (ECAP) during 2025 in the group of Dr. Alison Mitchell. HPHN also acknowledges the 2022 Theoretical and Computational Astrophysics Summer Student Program of the National Center for Theoretical Sciences (NCTS), Taiwan, where this project was initiated. 
ERO is supported by the RIKEN Special Postdoctoral Researcher Program for junior scientists, the RIKEN Incentive Research Project ("Diffusion and beyond: probing anomalous cosmic-ray transport in magnetized
molecular clouds"), and also acknowledges support from a Postdoctoral Fellowship of the Japan Society for the Promotion of Science (JSPS KAKENHI Grant No. JP22F22327), hosted by the University of Osaka, during which part of this work was completed. The early stages of this project were initiated when ERO was at the Center for Informatics and Computation in Astronomy (CICA). CICA is hosted by National Tsing Hua University and supported by a grant from the Ministry of Education of Taiwan. 
NT acknowledges support by JSPS KAKENHI grant Nos. JP22K14064 and JP26H00820. 
STC acknowledges support from the National Science and Technology Council (NSTC) of Taiwan under grants 112-2112-M-007-011, 113-2112-M-007-004, and 114-2112-M-007-001.
 
\begin{ack}
The authors thank the anonymous reviewer for their constructive comments, and
 Ruizhi Yang (University of Science and Technology of China) for providing reformatted \textit{Fermi}-LAT emissivity data of molecular clumps in Taurus, originally published in~\citet{Yang2023NatAs}. 
ERO is also grateful to Todd A. Thompson (The Ohio State University), Norman Murray (University of Toronto), Ellen G. Zweibel (UW--Madison), Hidetoshi Sano (Gifu University), and Sheng-Jun Lin (ASIAA) for discussions that helped shape the interpretation and consideration of the broader implications of this work. This study further benefited from discussions at the workshop \textit{Decoding Galactic Evolution through the Interplay of the Multi-Phase Interstellar Medium} (Nagoya, Japan, 2025). This research has made use of the {\tt SciPy} scientific computing package~\citep{2020SciPy-NMeth}, the {\tt naima} package for computation of non-thermal radiation from relativistic particle populations~\citep{naima}, the {\tt AAFragpy} package for hadronic interaction modeling~\citep{Koldobskiy2021PhRvD}, the {\tt matplotlib} library~\citep{Hunter:2007}, and NASA’s Astrophysics Data System. 
\end{ack}

\makeatletter
\let\PASJnormal@section\section
\let\PASJnormal@ssect\@ssect
\let\PASJnormal@sect\@sect
\makeatother

\appendix

\section{\textsc{Galprop} ISM Spectral Model}
\label{sec:appendix}

The interstellar CR electron and proton spectra for the boundary condition of each cloud was modeled using the \textsc{Galprop}~\citep{Vladimirov2011CoPhC}, 
covering an energy range up to 1 PeV. This choice ensured secondary X-ray synchrotron spectral shapes did not inherit cut-offs arising from the maximum energy choice in our boundary CR spectrum. 
The detailed parameters for our \textsc{Galprop} model are listed in Table~\ref{tab:GALPROP parameters}. 
\begin{table}
\caption{\textsc{Galprop} parameters were set as default values, 
expect for those listed below. Diffusive re-acceleration or convection effects were not considered. This is because simple diffusion prescriptions have been shown to be more consistent with synchrotron data and secondary-to-proton ratio constraints than diffusive re-acceleration models~\citep{Strong2011A&A}.  The propagation of primary electrons, secondary electrons, secondary positrons, secondary protons, secondary antiprotons, and tertiary antiprotons were considered and the outputs summed into electrons (all leptonic species) and protons (all hadronic species) to avoid unnecessary complication.}
\begin{threeparttable}
    {\centering
    \begin{tabularx}{\linewidth}{llc}
    \midrule
       \textbf{Parameters}  & \textbf{Definition} & \textbf{Value} \\
       \midrule
        \midrule
        $r_{\rm{max}}$ & Galaxy radius (kpc) & 25.0 \\
        $z_{\rm{max}}$ & Halo height (kpc) & 4.0\\ 
        $\Delta r$ & Radial step size (kpc) & 1.0 \\
        $\Delta z$ & Vertical step size (kpc) & 0.2 \\
        $D_{0, xx}$ & Diffusion coefficient\tnote{a}& $3.0 \times 10^{28}$ \\
        $R_{\rm{p}, \rm{br}}$ & Reference rigidity (GV) & 1.0 \\
        $\delta$ & Diffusion coefficient index\tnote{b} & 0.33 \\
        \midrule
        \multicolumn{3}{l}{\textit{Injected CR proton spectrum}}\\
        \midrule
        $R_{\rm{ref, p}}$ & Proton reference rigidity (GV)& 9.0 \\
        $\gamma_{\mathrm{p}, 1}$ & Injection index below $R_{\mathrm{ref, p}}$ & 1.7 \\
        $\gamma_{\mathrm{p}, 2}$ & Injection index above $R_{\mathrm{ref, p}}$ & 2.7 \\
        \midrule
        \multicolumn{3}{l}{\textit{Injected CR electron spectrum}} \\
        \midrule
        $\gamma_{\mathrm{e}, 0}$ & Injection index below $R_{\mathrm{e}, 1}$ & 1.60 \\
        $R_{\mathrm{e}, 1}$ & First rigidity break (GV) & 4.0 \\
        $\gamma_{\mathrm{e}, 1}$ & Index between  $R_{\mathrm{e}, 1}$ $\&$ $R_{\mathrm{e}, 2}$& 2.50 \\
        $R_{\mathrm{e}, 2}$ & Second rigidity break (GV) & 50.0 \\
        $\gamma_{\mathrm{e}, 2}$ & Injection index above $R_{\mathrm{e}, 2}$ & 2.70\\
        \midrule
    \end{tabularx}} 
\begin{tablenotes}
\footnotesize
\item[a] $D_{\mathrm{ISM}} = \beta D_{0, xx} (R/R_{\rm{p}, \rm{br}})^{\delta} \, \rm{cm}^2 \rm{s}^{-1}$ for rigidity $R$, $\beta = v/c$. 
\item[b] A diffusion coefficient index of 0.33, correspond to Kolmogorov turbulence, is adopted throughout this work. 
\end{tablenotes}
\end{threeparttable}
    \label{tab:GALPROP parameters}
\end{table}

\section{Fitting procedure for illustrative application}
\label{sec:fitting_procedure}

Below, we set out the detailed procedure used to conduct fits to obtain illustrative model parameter values for the Taurus molecular cloud and clumps. 
The procedure differs in the clumps compared to the overall cloud. This is because the cloud adopts the \textsc{Galprop} LIS as an outer boundary condition. For clumps, a functional form for the boundary is also fitted. 

\subsection{Full cloud}
\label{sec:fit_full_cloud}

The observational inputs are binned flux measurements over finite energy intervals $[E_{\mathrm{l},i},E_{\mathrm{h},i}]$. 
We convert the \citet{Neronov2017A&A} flux measurements to a 
spectral energy representation $F_i \equiv E_i^2\,\mathrm{d}N/\mathrm{d}E$ evaluated at the bin geometric mean $E_i=\sqrt{E_{\mathrm{l},i}E_{\mathrm{h},i}}$, with corresponding uncertainties. This is then converted to an aperture-matched per-H emissivity using a purely geometric normalization, 
\begin{equation}
q_{\gamma,\mathrm{obs}}(E) \;=\; F_{\mathrm{obs}}(E)\,\frac{4\pi d^2}{\cal N_{\rm{tot, H}}(<R_{\rm A})},
\end{equation}
where $d$ is the source distance and $\cal N_{\rm{tot, H}}(<R_{\rm A})$ is  
the total number of H atoms within the aperture, integrated from our adopted density profile (Equation~\ref{eq:n_gas}). 
This allows a direct comparison between the model-predicted per-H emissivity and the data-derived per-H emissivity for a similar region.

CR transport is governed by the model parameters introduced in section~\ref{sec:model}. These are $D_0$, with fixed energy scaling $D\propto D_0 E^{\delta}$, and $\eta$ which controls the effective scattering/penetration into dense gas in our model. To extract consistent estimations of these values from the $\gamma$-ray spectrum in terms of our model framework, 
we adopt a two-pass fitting strategy that reduces degeneracies between the boundary normalization parameters and the transport parameters. In the first stage (Stage 1) we fix $\eta=0.5$\footnote{This is simply chosen as the middle of the allowed range for this parameter. The results are not sensitive to the exact choice, unless $\eta$ is fixed close to the parameter bounds (0, 1) in Stage 1.} and fit $(\log_{10}D_0, A)$ over bounded ranges using a multi-start strategy. At each Stage-1 evaluation, the amplitude $A$ is obtained by best-fitting analytically in log-space via a weighted regression that accounts for asymmetric uncertainties. This reduces the dimensionality of the problem and improves numerical stability, and ensures $A$ is not treated as a free optimizer parameter. In the second stage of the fitting procedure (Stage-2) we freeze the amplitude adjustment, $A$, at its Stage-1 best value, and then fit $(\log_{10}D_0,\eta)$ using bounded optimization with multiple starting seeds in $\eta$. We include a soft Gaussian prior on $\log_{10}D_0$ centered on the Stage-1 optimum (with width specified in dex), to avoid large excursions while allowing transport parameters to adjust. The resulting best-fit values are reported in Table~\ref{tab:fit_results}, where we additionally show a measure for the 
goodness-of-fit in each case. This estimator is evaluated using a $\chi^2$-like statistic defined in logarithmic space,
\begin{equation}
\chi^2_{\log} \;=\; \sum_i \left[\frac{\log_{10}F_i-\log_{10}M_i}{\sigma_{\log,i}}\right]^2,
\end{equation}
where $M_i$ is the bin-averaged model SED and $\sigma_{\log,i}$ is obtained from the reported (possibly asymmetric) flux uncertainties, combined in quadrature with an additional fractional ``floor'' term that increases slightly with energy to prevent a small number of high-S/N points from dominating the fit. We report $\chi^2_{\log}/\nu$, where $\nu=N-k$ is the number of degrees of freedom given $N$ fitted SED points and $k$ fitted parameters. As $\nu$ is typically small in all our fits, we additionally quote the RMS of $\Delta\log_{10}F$ across the fitted points as an indicative dispersion metric. 

\subsection{Clumps}
\label{sec:fit_clumps}

In the case of conducting model fits for the clump sample, we modify our two-pass fitting strategy developed for the cloud to also simultaneously fit the boundary-shape parameters alongside the transport parameters. In the first stage (Stage 1) we again fix $\eta=0.5$ and fit $(\log_{10}D_0,a_2,\log_{10}E_c)$ over bounded ranges using a multi-start strategy. Clump-scale SEDs are noise-limited, so the adopted boundary shape can absorb spectral curvature. In our framework, this curvature is physically associated with the transport/attenuation effects we wish to capture in our fitting procedure. To prevent the boundary spectrum from artificially absorbing these signatures of transport suppression, we regularize the Stage-1 boundary parameters with Gaussian priors on $(a_2,E_c)$ anchored to the envelope spectrum reported by \citet{Yang2023NatAs}, i.e.\ $a_2=3.01\pm0.06$ and $E_c=2.71\pm0.65~\mathrm{GeV}$. In practice these are implemented as quadratic penalty terms added to the log-space objective. 

In the second stage of the fitting procedure (Stage-2) we freeze the boundary-shape parameters $(a_2,E_c)$ and (as previously) the amplitude $A$ at their Stage-1 best values, and then fit $(\log_{10}D_0,\eta)$ using bounded optimization with multiple starting seeds in $\eta$. We include a soft Gaussian prior on $\log_{10}D_0$ centered on the Stage-1 optimum (with width specified in dex), to avoid large excursions while allowing transport parameters to adjust. 
For the results shown here we use broad multi-start grids in $(\log_{10}D_0,a_2,\log_{10}E_c)$ and $\eta$ to reduce sensitivity to local minima. We again conduct the fit over a similar physical aperture as the data, $R_{\rm A}$, which 
we consider to be comparable to the smallest scale observable by \textit{Fermi}-LAT. At the distance of Taurus, this is typically around 0.25 pc (see Table~\ref{tab:fiducial_model} for exact parameter values). The size of the clumps and location of the boundary condition is typically larger than this, meaning that the $\gamma$-ray emission would be more sensitive to differences in transport physics. Our fiducial model results (see Figures~\ref{fig:ionization_results} and~\ref{fig:gamma_spec_clump}) demonstrated that 
effects are clearest in the central regions of clumps. The $\gamma$-ray emission volume of a cloud or clump 
is generally dominated by the outer layers, where effects are less distinct. Therefore, focusing the analysis window 
on the region where the strongest effects arise allows for more 
sensitivity to the CR transport physics that we intend to probe. 

\section{X-ray sensitivities}
\label{app:xray_sens}

The dotted sensitivity curves in Figure~\ref{fig:xray_predictions_taurus} are intended as order-of-magnitude, counts-limited surface-brightness scalings (not full detectability forecasts). We estimate them from published/anticipated on-axis effective areas $A_{\rm eff}(E)$ by requiring a minimum of $N_{\rm{src}}^{\min}=50$ source counts in one logarithmic energy bin ($\Delta\log_{10}E=0.2$) for an exposure time of 10 ks. For a source with surface brightness $I(E)$ extracted over solid angle $\Omega$, the expected counts in a bin are $N_{\rm{src}} \simeq I(E)\,A_{\rm eff}(E)\,T_{\rm exp}\,\Omega\,\Delta E$, and we therefore plot $I_{\min}(E) \propto N_{\rm{src}}^{\min}/[A_{\rm eff}(E)\,T_{\rm exp}\,\Omega\,\Delta E]$. To avoid unphysical sensitivity gains for sources larger than the usable field of view, we cap $\Omega$ by the effective single-pointing field-of-view area of each instrument. This simplified scaling ignores instrumental backgrounds, vignetting, PSF losses, and spectral redistribution.

Instrument inputs are taken from the \textit{Chandra}/ACIS-I Proposers' Observatory Guide (\url{https://cxc.harvard.edu/proposer/POG/html/index.html}, accessed January 2026), the \textit{XMM-Newton}/EPIC-pn Users Handbook~\citep{XMMUHB}, 
the AXIS mission page (\url{https://axis.umd.edu/researchers/mission}, accessed January 2026; see also~\citealt{Reynolds2023SPIE12678E}), and \textit{Athena}/WFI~\citep{Cruise2025NatAs}.

\makeatletter
\begingroup
\providecommand{\refname}{References}

\def\PASJrefheading#1{%
  \par\addvspace{1.5\baselineskip}%
  \noindent{\large\sffamily\bfseries #1}\par\nobreak
  \addvspace{0.5\baselineskip}%
}

\providecommand{\bibsection}{}
\renewcommand{\bibsection}{\PASJrefheading{\refname}}

\let\PASJorigsection\section
\def\section{\@ifstar{\PASJrefheading}{\PASJorigsection}}

\bibliographystyle{pasj} 
\bibliography{references} 
\endgroup

\end{document}